\documentclass[useAMS,usenatbib]{mnras}
\usepackage{graphicx}
\usepackage[utf8x]{inputenc}
\usepackage{journals}

\newcommand{\HII}{H{\sc ii}}

\begin{document}

\title[vdB 130 and interstellar extinction toward the Cyg OB1 supershell]{A revision of the vdB 130 cluster stellar content based on GAIA DR2 Data. Interstellar extinction toward the Cyg OB1 supershell.}
\author[Sitnik et al.]{
   T.G.~Sitnik$^{1}$\thanks{E-mail:tat.sitnik2011@yandex.ru},
   A.S.~Rastorguev$^{1,2}$\thanks{E-mail:alex.rastorguev@gmail.com},
   A.A.~Tatarnikova$^{1}$,
   A.M.~Tatarnikov$^{1}$,\newauthor
   O.V.~Egorov$^{1}$,
   A.A.~Tatarnikov$^{2}$\\
 $^{1}$Lomonosov Moscow State University, Sternberg Astronomical Institute,
        Universitetsky pr. 13, Moscow 119234, Russia \\
 $^{2}$ Lomonosov Moscow State University, Faculty of physics, Leninskie Gory 1 b.2, Moscow 119992, Russia}

\date{Accepted 2020 August 28. Received 2020 August 07; in original
form 2020 May 02}

\pagerange{\pageref{firstpage}--\pageref{lastpage}} \pubyear{2020}

\maketitle

\label{firstpage}

\begin{abstract}

Two star-forming regions are studied: the young embedded open cluster vdB 130
and the protocluster neighbourhood observed in the head and tail of the
cometary molecular cloud located in the wall of the expanding supershell
surrounding the Cyg OB1 association. The GAIA~DR2 catalogue is employed to
verify the stellar composition of the vdB~130 cluster whose members were
earlier selected using the UCAC4 catalogue. The new sample of vdB 130 members
contains 68 stars with close proper motions (within 1 mas yr$^{-1}$) and
close trigonometric parallaxes (ranging from 0.50 to 0.70 mas). The relative
parallax error is shown to increase with distance to objects and depend on
their magnitude. At a distance of 1.5--2~kpc it is of about 3--7 per cent and
20--30 per cent for bright and faint stars, respectively. The cluster
is not older than $\sim$10 Myr. New spectroscopic and photometric
observations carried out on Russian telescopes are combined with GAIA~DR2 to
search for optical components in the protocluster region -- a new starburst.
An analysis of 20 stars in the vicinity of the protocluster revealed no
concentration of either proper motions or parallaxes. According to
spectroscopic, photometric, and trigonometric estimates, the distances to
these stars range from 0.4 to 2.5 kpc, and colour excess is shown to increase
with a distance $D$ (kpc) in accordance with the law: $E(B-V)\simeq 0.6
\times D$~mag.

\end{abstract}


\section{Introduction}

Nowadays it is well established that star formation could be regulated by many factors like turbulence, feedback from outflows, supernova explosions and
expanding \HII\ regions \citep[e.g., see review by][]{Elmegreen1998, Elmegreen2010}. These and other factors work together sculpting a
fine structure of molecular clouds with filaments, pillars, blobs, peppered
with young stellar objects (YSOs) in the rims of expanding bubbles. A growing number of observational works supports the idea of sequential and triggering star formation on  both local and global scales \citep[e.g.,][among many others]{zavagno2006, deharveng2010, Dale2015, Egorov2017}. Given complexity of the process, along with
large-scale studies, a detailed look at some specific objects and regions can
be useful for distinguishing between general and particular features of the
star formation process.

The aim of this work is to continue the study of star formation
regions in the walls of an expanding supershell formed by the wind and the UV
radiation of stars in the Cyg OB1 association. We present an investigation of
the stellar content of the cluster vdB 130 and of optical counterparts of the
protocluster using the new catalogue GAIA~DR2 (\citealt{GDR2}\footnote{VizieR On-line
Data catalogue:I/345/gaia2})

\begin{figure}
\includegraphics[width=\linewidth]{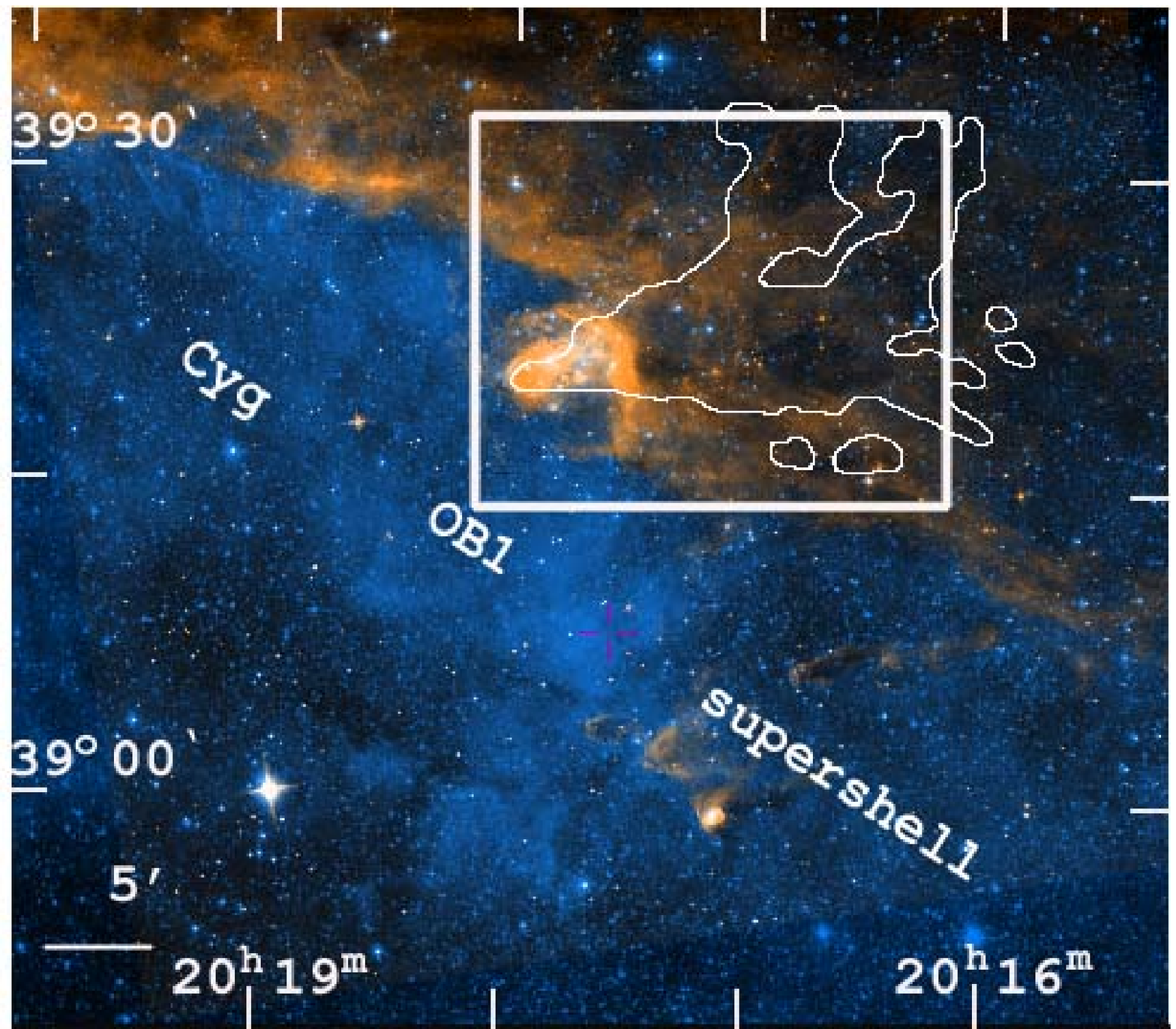}
\includegraphics[width=\linewidth]{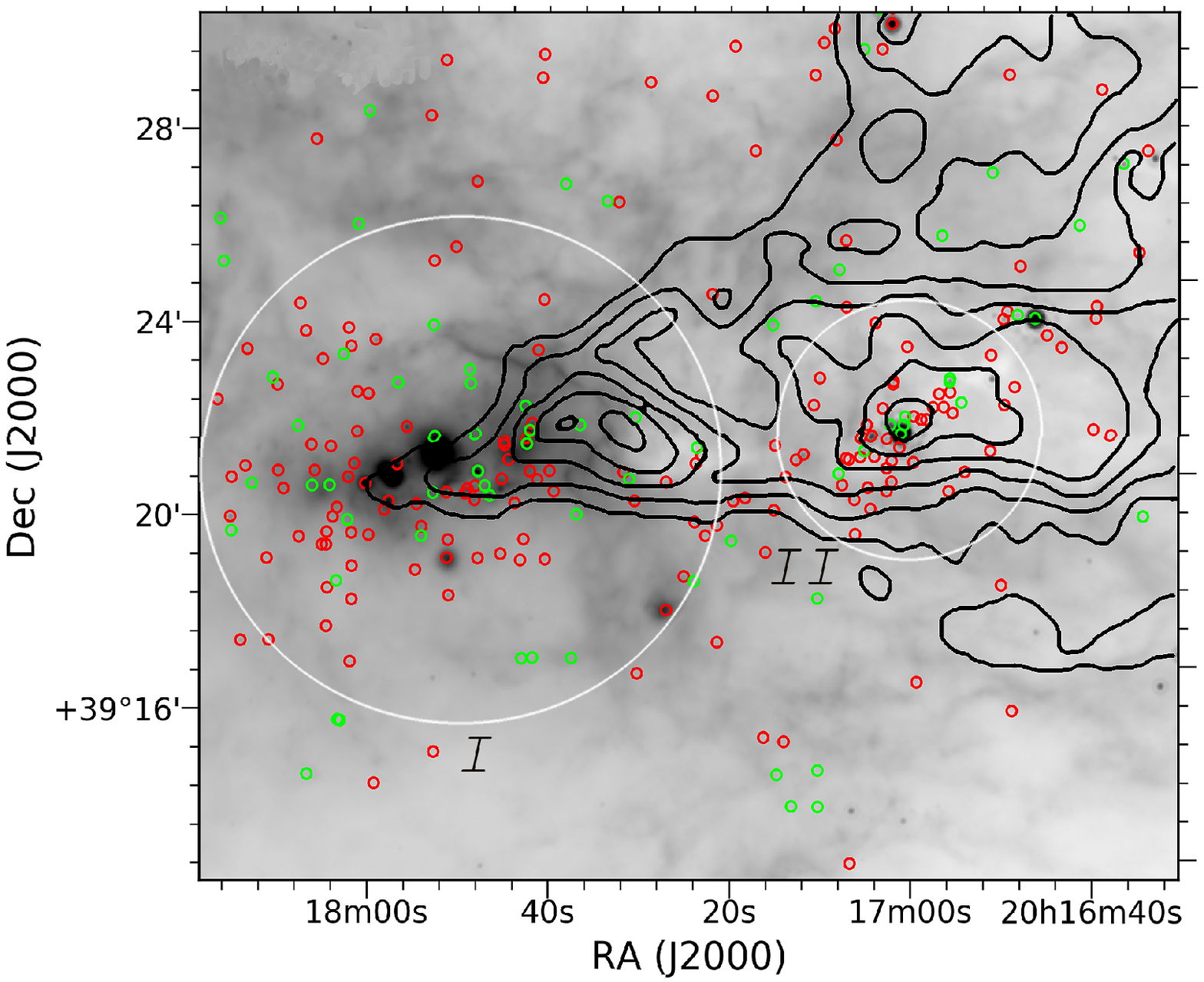}
\caption{Northwestern part of the supershell surrounding the stellar
association Cyg OB1. Top: superimposed images of the optical (blue) and IR
(red) maps (the latter is based on 5.8-$\mu$m \textit{Spitzer} data) and the
contour delimiting the molecular cloud~\citep{sch07}. Bottom: zoomed-in
24-$\mu$m image of the region in the rectangle shown in the top panel. The
circles mark the cluster vdB130 (I) and protocluster (II) regions. The green
and red circles indicate type I and II protostars, respectively (see Paper
II). Isocontours describe the molecular cloud mentioned above.}
\label{fig:cloud-proto}
\end{figure}

We have already partially studied the region of ongoing
star formation in the supershell related to the cometary molecular cloud
(pillar)(\citealt{sit15, sit19}, hereinafter referred to as Paper I and Paper
II, and \citealt{tatar16}). This region ($\alpha = 20^{h}16^{m} -
20^{h}18^{m}, \delta=39^{\circ}15^{'} - 39^{\circ}30^{'}$) resides in the
wall of the expanding supershell surrounding the Cyg OB1 stellar association
(Fig.\ref{fig:cloud-proto},top). The Cyg OB1 association includes at least 50
OB stars \citep{Humpreys}. The size of the supershell driven by this
association is $3^{o}\times4^{o}$. In the supershell region mentioned above
a cometary molecular cloud is observed~\citep{sch07}. It is associated with
two sites of star formation: the young embedded cluster~vdB 130 in the head
and a compact protocluster in the tail (Fig.\ref{fig:cloud-proto}, bottom).
We found this protocluster in a dense condensation of the cloud and
described it in Paper II (see  fig.~11 in Paper II). The cloud is extended
toward the illuminating source, i.e., toward the nearest OB stars of the Cyg
OB1 association. The molecular cloud has a size of 0.4~degrees or 12~pc at
the adopted distance of $1.8 \pm 0.3$ kpc to the cluster and association. The
vdB 130 cluster has 44 members selected based on UCAC4 proper motions. They
are listed in table~2 in Paper~I. The physical connection of the  Cyg OB1
association, supershell, and vdB 130 cluster, and the cloud follows from the
distance estimates, measured radial velocities, and a number of indirect
factors  (Paper I). The age of vdB~130 was estimated as 5--10~Myr. According
to near-IR data, the true distance modulus to the cluster, its minimum colour
excess, and distance are equal to $(K-M_ks)=11.26\pm 0.30$~mag,
$E(J-H)=0.27\pm0.02$~mag, and about 1.8 kpc, respectively. The interstellar
extinction law inside the cluster region differs significantly from the
normal law, with $R_{v}=A_{v}/E(B-V)$ reaching values of $6-8$. On the other
hand, the normal law is valid in front of the cluster \citep{tatar16}. Some
cluster stars are located in the centres of dust clumps inside the IR
shell.

We use the data from GAIA~DR2 (\citealt{GDR2})%, VizieR On-line Data
%catalogue:I/345/gaia2)
to analyse the star membership in two groupings in the
supershell surrounding the Cyg OB1 association -- in  the cluster vdB~130 and
in the protocluster. The embedded cluster vdB 130 was initially identified by
\citet{rac74} as a group of 14 B-type stars. We added another 30 stars with
close proper motions from the UCAC4 catalogue (in terms of the accuracy of
the UCAC4 catalogue) (Paper I). However, the GAIA DR2 catalogue allows us to
appreciably refine the selection of stars both by proper motions and
distances. It is especially important for objects observed toward the Cygnus
arm, because the proper motions due to the differential rotation of the
Galaxy vary slightly with distance.

We also performed optical and infrared photometric and spectroscopic
observations on Russian telescopes in order to clarify the spectral
classification of stars, study the spectral energy distribution, and
estimated the value of interstellar extinction.

In this paper we use new observational data, as well as archival data from
the GAIA DR2 catalogue in order to verify the vdB~130 cluster earlier studied
using UCAC4 proper motions and search for possible optically visible member
stars of the protocluster. We use our data to study interstellar extinction
toward the supershell formed by Cyg OB1 stars. Section~2 describes the
instruments and data reduction process employed for spectroscopic and
photometric observations. Section~3 analyses the line-of-sight distribution
of vdB~130 stars and potential members of the optical component of the
protocluster (using GAIA DR2 catalogue). In the same section we
estimated the age of the cluster using stars with a known spectral
classification \citep{tatar16}. Section~4 presents the results of the
spectral classification of stars seen toward the protocluster and  colour
excess estimates. In section~5 interstellar extinction toward the cometary
molecular cloud is analysed using GAIA DR2 results among other data.
Section~6 summarizes the results of the study.

\section{Observations and data reduction.}

This study is based on optical and near-IR observations performed with
the 6-m telescope of Special Astrophysical Observatory of Russian
Academy of Sciences (SAO RAS), 2.5-m telescope of Caucasian Mountain Observatory of Sternberg
Astronomical Institute of M.V. Lomonosov Moscow State University (SAI MSU) and 60-cm telescope of the Crimean
Astronomical Station of SAI MSU. The archival data
obtained with \textit{Spitzer} and \textit{GAIA} (DR2) space
observatories were also used. Below we present a
description of the observations and data
reduction. The log of our observations is given in
Table~\ref{tab:obs_data}, where the exposure time, $\mathrm{T_{\rm
exp}}$, the field of view (FOV), the final angular resolution,
$\theta$, the final spectral resolution, $\delta\lambda$, and the
bandwidth (FWHM) of the used filter are indicated.

\begin{table}
\caption{Log of observations.}
\begin{scriptsize}
\label{tab:obs_data}
\begin{tabular}{lcllc}
\hline
Instrument &{Slit/filters (stars)}& Date    & $\mathrm{T_{exp}, s}$ & $\theta$,$''$ \\
\hline
SCORPIO     &  \#1 (6, 7, 12, 23) &  2017 Jul 28 & 180 &   1.8  \\ %(PA = 203\degr)   % 2017 Jul 28   2457963.22
SCORPIO     &   \#2 (2,9,17, 22) &2017 Jul 28   & 240 &   1.8 \\ %(PA = 226\degr)  % 2017 Jul 28   2457963.23
SCORPIO     &   \#3  (5, 13) & 2017 Jul 28   & 180 &    1.5\\ %(PA = 187\degr) % 2017 Jul 28   2457963.25
SCORPIO     &   \#4 (3, 8, 10)  & 2017 Jul 28  & 1320& 1.5 \\ %(PA = 267\degr) % 2017 Jul 28   2457963.27
SCORPIO     &   \#5 (15, 19, 21) & 2017 Jul 30  & 1800&  3.0\\ %(PA = 276\degr) % 2017 Jul 30  2457965.45
SCORPIO     &   \#6 (11, 18)  & 2017 Jul 31  & 1200& 3.0\\ %(PA = 20\degr) % 2017 Jul 31 2457966.44
SCORPIO     &   \#7  (13, 14, 20) & 2017 Jul 31  & 791 &  3.0 \\ % (PA = 50\degr) % 2017 Jul 31 2457966.46
ASTRONIRCAM &    $JHK$              & 2016 Sep 17  & 40  &  0.8    \\ % 2016 Sep 17
CCD photometer &    $BVR_c$         & 2018 Aug 05  & 600   &  1.2 \\ % 2018 Aug 05
\hline
\end{tabular}
\end{scriptsize}
\end{table}

\subsection{Spectroscopic observations and data reduction.}

\begin{figure}
\includegraphics[width=\linewidth]{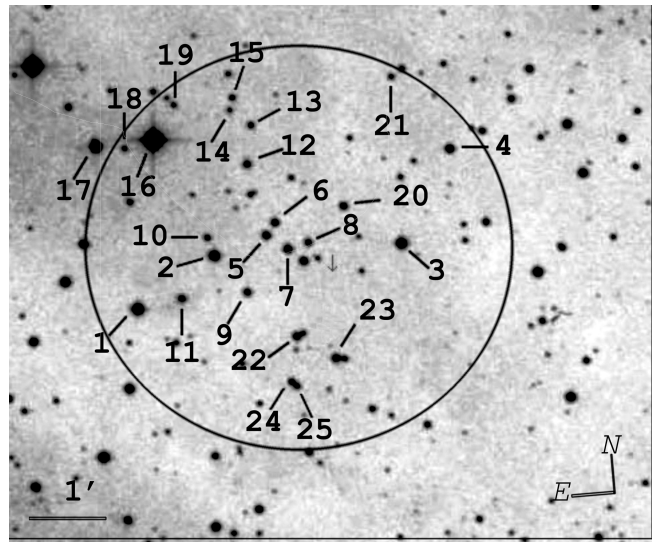}
\caption{The optical map of the area in the direction of protocluster. The
stars selected for spectroscopic study are indicated. The circle marks the
same area II, as in Fig.\ref{fig:cloud-proto},bottom}
\label{fig:cloud-proto1}
\end{figure}

Spectroscopic observations on the 6-m telescope of SAO RAS were performed
using SCORPIO multimode focal reducer~\citep{scorpio} operating in the
long-slit spectroscopy mode. We used the slit of the 1 arcsec width and 6.1 arcmin length. The scale along the slit was 0.36 arcsec per pixel.
The spectra were registered with  EEV 42-40 CCD using VPHG550G prism as the
dispersing element, which provided a spectroscopic resolution of
$\delta\lambda = 13$~\AA\, (measured as FWHM of the air-glow lines) in the
$3500-7500$~\AA\, wavelength interval. We acquired spectra of stars and of
the surrounding interstellar medium in the protocluster region at seven
different slit positions; the information about all the obtained data is
summarized in Table~\ref{tab:obs_data}. Each slit passed through 2--4 stars of our interest (their names according to Fig~\ref{fig:cloud-proto1} are given in parenthesis near the slit's number). The observations with first three slit positions were performed with short exposures because the stars crossed by the slit were bright.  We reduced the data using standard procedures realized in \textsc{idl} pipeline developed at the SAO RAS for
analysing the data acquired with SCORPIO spectrograph. The data reduction
included bias subtraction, flat-field and line-curvature correction,
wavelength calibration, and subtraction of night sky lines. A He-Ne-Ar lamp
served as the source of the comparison spectrum for deriving the dispersion
curve, and the spectrophotometric standard stars BD+25d4655 and BD+33d2642
observed at similar zenith angles before and after the object were used to
calibrate the spectra. Unfortunately,  spectrophotometric standard for slits
6 and 7 (stars 11, 13, 14, 18 and 20, see Fig~\ref{fig:cloud-proto1}) could not be observed because of
weather conditions during the last night. In current analysis we were
interested in relative flux calibration only, so for these spectra we used
the sensitivity curve based on the data obtained during the previous night.

We analysed  the spectra of stars observed in the neighbourhood of the
protocluster (Fig.\ref{fig:cloud-proto1}) in order to determine their
spectral types and, in some cases, also the luminosity classes . The
procedure of the determination of stellar parameters consisted of three
stages. At the first stage the spectral types of stars were estimated using
the methods described by \citet{gray}. We used continuum-normalized spectra
of MK standards from the same source. We divided the spectrum by a
fifth-degree polynomial describing the form of the continuum spectrum of the
object. Then we compared it with the spectra of MK standards. However, in the
case of late- (M and K) type stars, whose spectra exhibit strong molecular
bands, we compared the spectrum of the object with that of the standard
directly without normalizing the object spectrum to the continuum. Our
spectroscopic data do not allow the luminosity class to be estimated with
sufficient precision except for late-type stars. Therefore we assumed, to a
first approximation, that the stars studied are of luminosity class V except
for star 17 (Fig.\ref{fig:cloud-proto1}, see also Section~4), which can be
already classified as having luminosity class III. (We will later verify the
luminosity class using photometric data and the distances inferred from GAIA
DR2 trigonometric parallaxes.) The low spectral resolution and low
signal-to-noise ratio of our observations make it also impossible to study
the peculiarities of the chemical composition. We assumed in all cases that
the objects studied match standard stars with no chemical anomalies. The only
exceptions are stars 12 and 14 ( Fig.\ref{fig:cloud-proto1}, see also
Section~4). Thus the Ca~\textsc{ii} line in the spectrum of star 12 is
significantly weaker than expected for stars of the same spectral type,
whereas the spectrum of star 14 exhibits bright Eu~\textsc{ii} 6049.5~\AA\,
line (see below).

At the second stage we determined the colour excess $E(B-V)$ based on the
spectral type estimated at the previous stage and the observed colour index
$(B-V)_{obs}$. We adopt the intrinsic optical and IR colour indices
from~\citet{str} and used the intrinsic colour index $(J-K)_{0}$ to calculate
the $(J-K)_{calc}$ colour index assuming normal interstellar extinction law.
If all the assumptions are true then the calculated colour index
$(J-K)_{calc}$ must coincide with the observed colour index $(J-K)_{obs}$.

\begin{table*}
\caption{Photometric parameters of stars toward the protocluster. Column~1 -
number of star according to Fig.\ref{fig:cloud-proto1}, columns 2-7 -
brightness in bands $B$, $V$, $R_c$, $J$, $H$, $K$ (mag), column 8 -  the
trigonometric parallaxes of the stars (mas), column 9 - the corresponding
distances (kpc) (without he systematic correction), columns 10 and 11 - the
proper motion (mas\ yr$^{-1}$). Subscript 2 represent transformed 2MASS data.}
\label{tab:stars}
\begin{scriptsize}
\begin{tabular}{c|c|c|c|c|c|c|c|c|c|c}
\hline
Star \# & $B$, mag &$V$, mag &$R_c$, mag &$J$, mag & $H$, mag& $K$, mag &  plx, mas & D, kpc & $ \mu_{\alpha}$, mas/yr & $ \mu_{\delta}$, mas/yr \\
\hline
 1  &  $13.230 $ &  $12.691 $ & 12.418 &  11.445  &  11.268     &  11.209  &  $0.582\pm0.029$ &  $1.72\pm0.09$ & $-5.040\pm0.044$  & $-3.130\pm0.044$ \\
 2  &  $14.302 $ &  $13.574 $ & 13.201 &  12.118  &  11.896     &  11.762  &  $0.719\pm0.016$ &  $1.39\pm0.03$ & $-5.575\pm0.026$  &  $-12.665\pm0.025$\\
 3  &  $13.689 $ &  $13.181 $ & 12.940 &  12.203  &  $11.977_2$ &  11.939  &  $1.247\pm0.016$ &  $0.80\pm0.01$ & $-2.157\pm0.026$  &  $-4.524\pm0.027$ \\
 4  &  $15.598 $ &  $14.756 $ & 14.314 &  13.110  &  $12.693_2$ &  12.683  &  $0.958\pm0.023$ &  $1.04\pm0.03$ & $-2.544\pm0.037$  &  $-7.455\pm0.035$ \\
 5  &  $15.856 $ &  $15.069 $ & 14.676 &  13.562  &  13.199     &  13.111  &  $1.016\pm0.023$ &  $0.98\pm0.02$ & $-12.359\pm0.038$ & $-28.540\pm0.036$ \\
 6  &  $16.461 $ &  $15.470 $ & 14.884 &  13.500  &  12.973     &  12.839  &  $1.916\pm0.026$ &  $0.52\pm0.01$ & $13.611\pm0.044$  & $14.532\pm0.041$\\
 7  &  $15.467 $ &  $14.758 $ & 14.391 &  13.350  &  13.029     &  12.926  &  $1.259\pm0.029$ &  $0.79\pm0.02$ & $6.080\pm0.035$   &  $3.213\pm0.033$\\
 8  &  $16.880 $ &  $15.971 $ & 15.466 &  14.017  &  13.623     &  13.447  &  $0.749\pm0.036$ &  $1.34\pm0.06$ & $1.121\pm0.059$   &  $7.184\pm0.057$\\
 9  &  $16.484 $ &  $15.683 $ & 15.266 &  14.084  &  13.717     &  13.663  &  $0.867\pm0.030$ &  $1.15\pm0.04$ & $-3.847\pm0.051$  & $-15.789\pm0.047$\\
10  &  $17.515 $ &  $16.480 $ & 15.901 &  14.286  &  13.913     &  13.725  &  $0.541\pm0.041$ &  $1.85\pm0.14$ & $-2.898\pm0.069$  &  $-6.163\pm0.068$\\
11  &  $17.591 $ &  $16.269 $ & 15.427 &  12.356  &  11.440      &  10.804  &  $0.530\pm0.095$ &  $1.87\pm0.34$ & $-5.730\pm0.168$  &  $-4.115\pm0.164$\\
12  &  $16.652 $ &  $15.746 $ & 15.240 &  13.726  &  13.390      &  13.238  &  $0.601\pm0.031$ &  $1.66\pm0.09$ & $-0.743\pm0.050$ &  $-2.183\pm0.050$\\
13  &  $17.512 $ &  $16.456 $ & 15.920 &  14.481  &  14.093     &  13.966  &  $0.716\pm0.044$ &  $1.40\pm0.09$ & $-1.344\pm0.071$ &  $-4.614\pm0.078$\\
14  &  $19.6   $ &  $18.038 $ & 16.891 &  13.434  &  12.707     &  12.228  &  $0.364\pm0.072$ &  $2.75\pm0.54$ & $-2.279\pm0.117$ &  $-3.806\pm0.116$\\
15  &  $18.191 $ &  $16.802 $ & 15.966 &  13.905  &  13.262     &  13.020  &  $2.618\pm0.050$ &  $0.38\pm0.01$ & $-7.551\pm0.077$ & $-40.80\pm0.074$\\
16  &  $<11.3  $ &  $<10.8  $ & 10.304 &$8.153_2$ & $7.618_2$   &$7.518_2$ &  $2.387\pm0.031$ &  $0.42\pm0.01$ & $10.326\pm0.046$ &   $4.836\pm0.046$\\
17  &  $15.077 $ &  $13.326 $ & 12.364 &   9.849  & $9.039_2$   &$8.763_2$ &  $0.613\pm0.031$ &  $1.63\pm0.08$ & $-3.116\pm0.047$ & $-12.758\pm0.077$\\
18  &  $17.950 $ &  $16.826 $ & 16.216 &  14.661  &  14.210      &   14.042 &  $0.755\pm0.052$ &  $1.32\pm0.09$ &  $0.657\pm0.085$ &  $-1.120\pm0.085$\\
19  &  $18.730 $ &  $17.360 $ & 16.436 &  14.204  &  13.570      &   13.337 &  $2.680\pm0.057$ &  $0.37\pm0.01$ &  $2.133\pm0.092$ & $-10.853\pm0.097$\\
20  &  $16.514 $ &  $15.613 $ & 15.142 &  13.787  &  13.341     &   13.176 &  $1.449\pm0.075$ &  $0.69\pm0.04$ &  $3.245\pm0.130$ &   $6.295\pm0.117$\\
21  &  $18.374 $ &  $17.156 $ & 16.530 &  14.854  & $14.374_2$  &   14.237 &  $0.668\pm0.065$ &  $1.50\pm0.15$ &  $0.804\pm0.103$ &  $-4.492\pm0.095$\\
22  &  $16.383 $ &  $15.415 $ & 14.893 &  13.414  & $13.026_2$  &   12.949 &  $0.654\pm0.025$ &  $1.53\pm0.06$ & $-7.981\pm0.042$  &  $-7.181\pm0.039$\\
23  &  $15.724 $ &  $14.999 $ & 14.627 &  13.536  & $13.224_2$  &   13.183 &  $0.831\pm0.025$ &  $1.20\pm0.04$ &  $1.098\pm0.040$ &  $-2.532\pm0.036$\\
24  &  $17.176 $ &  $16.176 $ & 15.633 &  14.166  & $13.738_2$  &   13.710 &  $0.713\pm0.035$ &  $1.40\pm0.07$ & $-2.782\pm0.059$  &  $-1.081\pm0.056$\\
25  &  $17.355 $ &  $16.411 $ & 15.587 &  14.461  & $14.044_2$  &   14.038 &  $0.604\pm0.040$ &  $1.66\pm0.11$ & $-3.308\pm0.067$  &  $-0.961\pm0.063$\\
\hline
\end{tabular}
\end{scriptsize}
\end{table*}

\subsection{Photometric observations.}

We carried out \textit{BVRc} photometry observations of 25 stars in the
protocluster region (Fig.\ref{fig:cloud-proto1}) with Apogee Aspen CG-42
camera (image scale 0.37 arcsec pix$^{-1}$, field of view 12.6 arcmin) attached to the 60-cm telescope
of the Crimean Astronomical Station of SAI MSU. The standards employed were
11--15 mag field stars with photometry errors smaller than 0.05 mag. We
adopted the \textit{B, V, g', r'}, and \textit{i'}-band magnitudes of these
stars from the APASS catalogue\footnote{VizieR On-line Data catalogue:II/336} and used
the colour equations from \citet{Jordi} to transform APASS data to the
$R_c$-band filter. We performed the relative photometry of all selected
standard stars with respect to each other. The accuracy of the resulting
mutual calibration is better than 0.01 mag.

To obtain $JHK$-band data in the Mauna Kea Observatory (MKO) system, we
observed our stars with ASTRONIRCAM IR camera of the 2.5-m telescope of the
Caucasian Mountain Observatory of SAI MSU \citep{Nadjip17}. The field of
view of the camera has a size of 4.6~arcmin and the image scale is 0.27
arcsec pix$^{-1}$. The standards employed were field stars fainter than
10~mag whose 2MASS magnitudes have errors smaller than 0.05~mag. We
transformed the 2MASS data to MKO magnitudes using the colour equations
from~\citet{Leggett}. Observations were performed using the dithering method,
with the telescope shifting between individual frames by $3\times4$ arcsec.
Each frame was corrected for non-linearity, dark current and flat field. We
reduced both optical and IR frames using aperture photometry technique and
standard astrometry.net \citep{Lang} and
photutils\footnote{\url{https://photutils.readthedocs.io/en/stable/}}
software suites. We chose the aperture size to be $\sim2\times$FWHM ($\sim2$
arcsec). In Table~\ref{tab:stars} (columns  2--7) we present the results of
our photometry of the stars marked in Fig.\ref{fig:cloud-proto1}. The
accuracy of $BVR_c$-band photometry is better than 0.03 mag for stars
brighter than 17.5 mag with the error amounting to 0.1 mag for 19-mag stars.
The accuracy of $JHK$-band photometry is better than 0.02 mag. Stars 16 and
17 (see Section~4) proved to be too bright for ASTRONIRCAM, and we therefore
use transformed 2MASS data for these objects. We could not acquire $H$-band
photometry for some of the stars. In these cases we also used transformed
2MASS data (marked by subscript index `2' in the Table~\ref{tab:stars}).
Columns~8--11 give the trigonometric parallaxes of the stars, the
corresponding distances, and the proper motions of stars according to GAIA
DR2 data.

\subsection{Archival data}

In this study we also use  \textit{G}, \textit{BP} and \textit{RP} band
photometric data, colour excesses $E(BP-RP)$, trigonometric parallaxes, and
proper motions from  GAIA DR2 catalogue. We also applied the systematic
correction to trigonometric parallaxes, which we assumed to be equal to
+45~microarcseconds as estimated, e.g., from open
clusters~(\citealt{yalyalieva18}; see also \citealt{Groenewegen}). The exact
value of this correction is not very important for the distances considered,
1.5--2~kpc: its relative contribution does not exceed 8--10 per cent.

\section{Analysis of GAIA DR2  data for stars seen toward the molecular cloud.}

We use the data of GAIA DR2 catalogue to study optical objects in
star-forming regions located in the cometary molecular cloud, namely, in the
region of vdB 130 cluster and the protocluster (Fig.\ref{fig:cloud-proto},
bottom).

\subsection{The open cluster vdB 130.}

The vdB 130 cluster was initially identified by ~\citet{rac68,rac74} as a
compact group of B-type stars with a size of about~6 arcmin. In Paper I we
analysed UCAC4 proper motions of stars in this region and found that the
proper motions of 44 of 175 stars located inside the 6-arcmin radius region
centred on $\alpha_{2000}\sim 20^{\mathrm{h}}17^{\mathrm{m}}48^{\mathrm{s}}$,
$\delta_{2000}\sim 39^{\circ} 21^{\prime}00^{\prime\prime}$ differ by less
than 6 mas\ yr$^{-1}$ (see fig. 3 and table~2 in Paper~I). However, recall that vdB
130 is located in the Cygnus arm and is observed in the direction of Galactic
longitude of $\emph{l}\sim 77^{o}$. Therefore the stellar proper motions due
to differential rotation of the Galaxy vary slightly with distance. That is
why we decided to identify cluster members based on the proximity of not only
the GAIA DR2 proper motions but also of GAIA DR2 trigonometric parallaxes.

\begin{figure}
\includegraphics[width=0.92\linewidth]{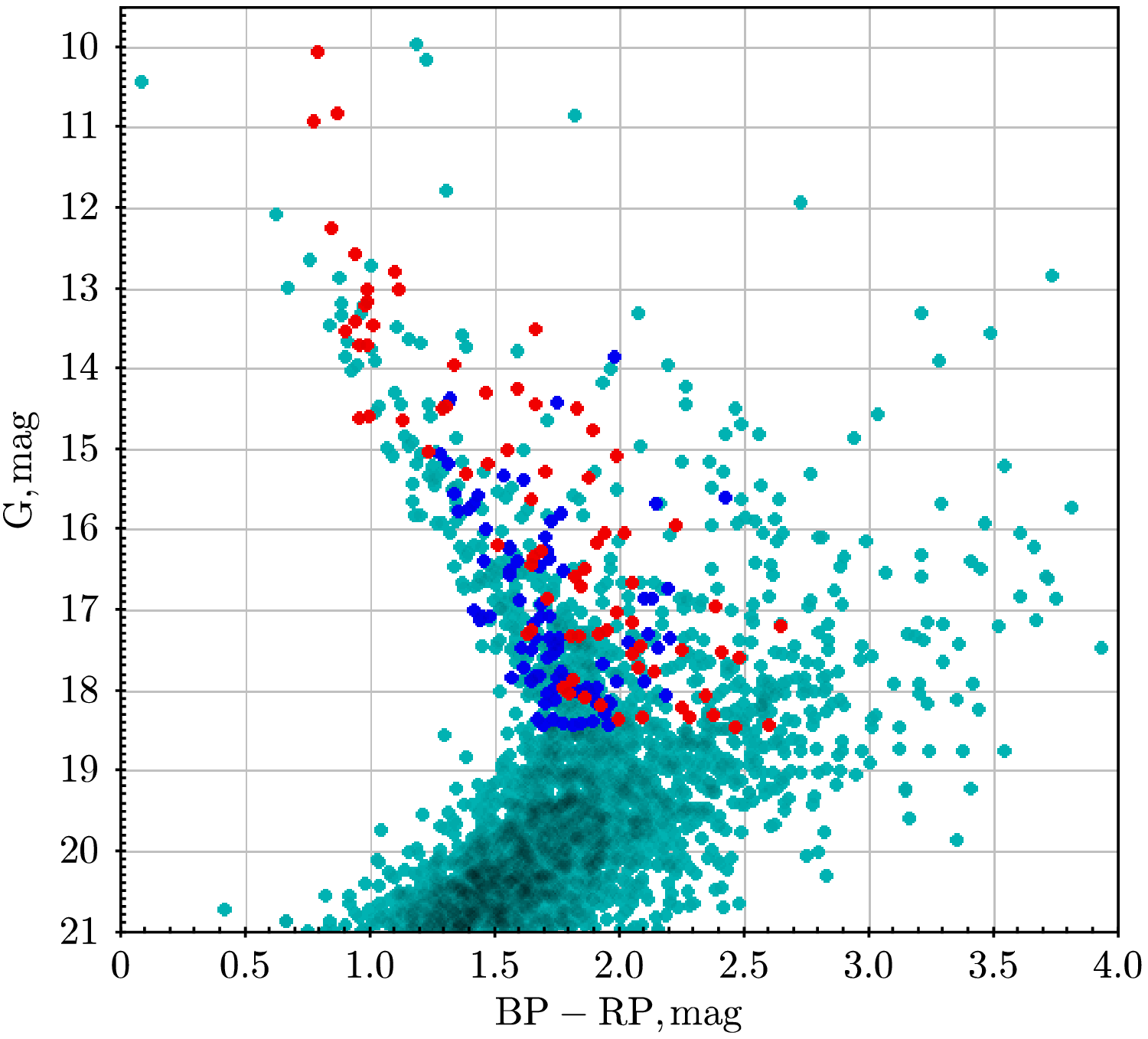}

\includegraphics[width=0.92\linewidth]{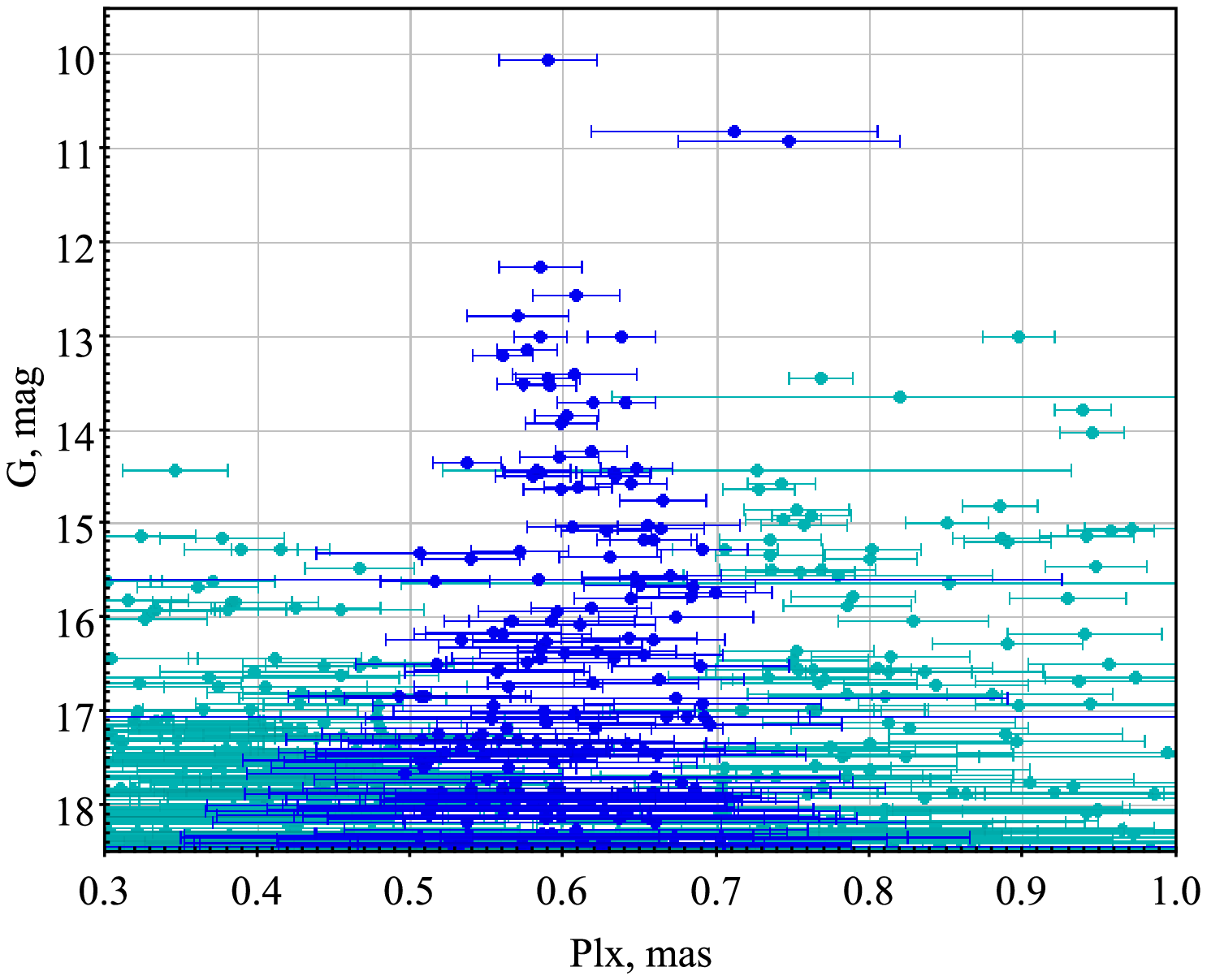}

\includegraphics[width=0.92\linewidth]{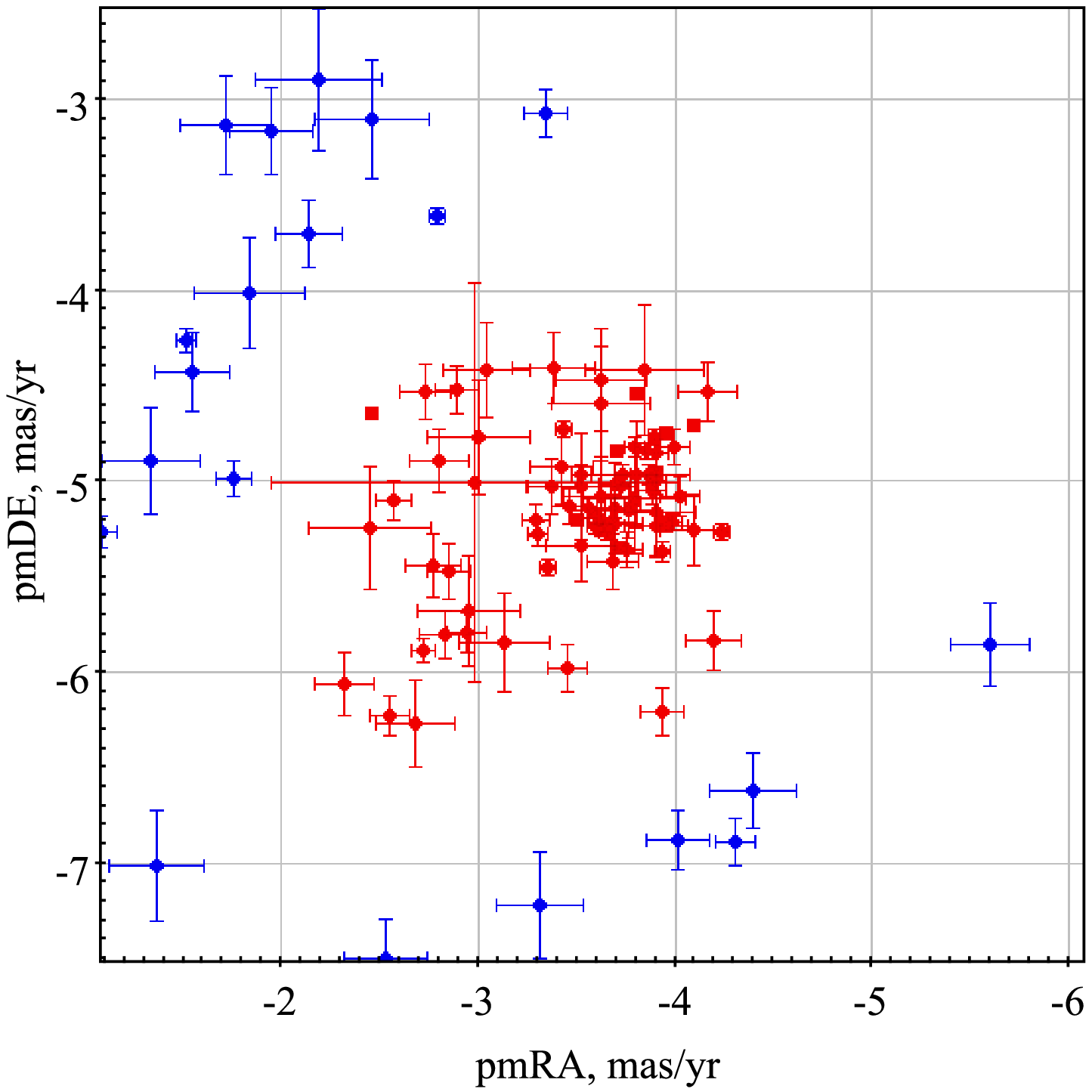}
\caption{Top panel: $(BP-RP)$ -- $G$ colour-magnitude diagram for all stars
in the direction of the vdB~130 cluster. Central panel: GAIA DR2 parallax
(with systematic correction of +0.045~mas) -- $G$ magnitude diagram. Stars
with DR2 parallaxes from 0.50 to 0.70~ mas are marked by blue. Bottom panel:
two-dimensional proper motion diagram for stars with close parallaxes.
Probable cluster members selected both by parallaxes and proper motions are
marked by red. Both selections are also marked on top panel.}
\label{fig:vdB130}
\end{figure}

\begin{figure*}
\includegraphics[width=\linewidth]{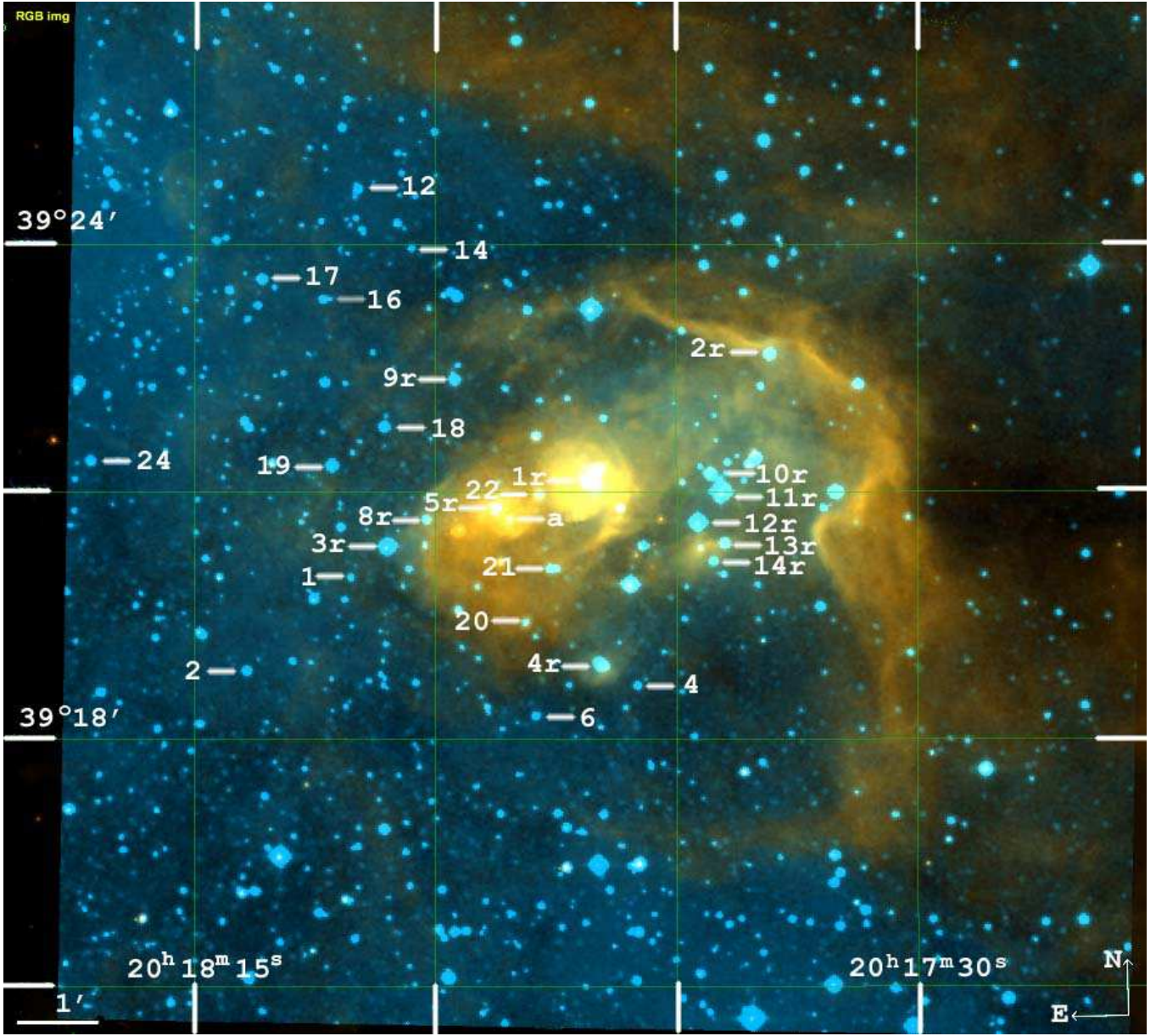}
\caption{Superimposed images of the blue optical (F-DSS) and red IR (8~$\mu$m
\textit{Spitzer}) maps of the area toward the vdB 130 cluster. Also shown are
the numbers of those cluster stars which have been selected from both
catalogues UCAC4 and GAIA DR2 (see section 3.1 for details). The stars that
initially identified by \citet{rac74} are marked by the letter 'r'.}
\label{fig:stars}
\end{figure*}

Top panel in Fig.~\ref{fig:vdB130} show the CMD \emph{(BP-RP) -- G} for all
stars in the 6-arcmin cone region (see Paper I) centred on $\alpha_{2000}\sim
20^{\mathrm{h}}17^{\mathrm{m}}51^{\mathrm{s}}$, $\delta_{2000}\sim 39^{\circ}
21^{\prime}06^{\prime\prime}$. In this region there are 3370 GAIA DR2 stars.
However, stars of the Cyg OB1 association also project onto this area.
Central panel on Fig.~\ref{fig:vdB130} shows GAIA DR2 parallaxes vs. G-band
magnitudes for all stars observed in this area. The stars in the narrow
parallax interval from 0.5 to 0.7 mas (i.e., with distances ranging from 1.5
to 2.0~kpc) are marked by dark blue dots, with appropriate error bars. The
bottom panel in Fig.~\ref{fig:vdB130} shows the two-dimensional proper-motion
diagram for selected stars with close parallaxes. Probable cluster members
with close parallax and proper motion values are marked by red dots; proper
motions errors are also shown. Note that the proper motions differ by no more
than 1 mas~yr$^{-1}$ from the centroid of proper motions. Probable cluster
members and stars with close parallaxes are also shown on top panel of
Fig.~\ref{fig:vdB130}.

The parallax interval selected above corresponds to the distance
interval from 1.5 to 2~kpc, which agrees well with the IR-data
based cluster distance estimate reported in Paper~I. In
Fig.~\ref{fig:vdB130}, bottom concentration of proper motions to
the centroid with ${(pmRA, pmDEC)} = (-3.70 \pm 0.35, -5.00 \pm
0.30)$ mas~yr$^{-1}$ is immediately apparent, and this agrees well
with our results based on UCAC4 catalogue (Paper~I) and on an
appreciably smaller number of stars.

The proper-motion centroid in the Galactic coordinate system is located at
$(pmL, pmB) = (-6.20 \pm 0.33, +0.20 \pm 0.33)$ mas~yr$^{-1}$. The cluster
obviously moves predominantly along the Galactic plane with residual velocity
of about $-8$~km~s$^{-1}$, i.e., toward the Galactic center (after
subtracting the contribution of differential rotation and solar motion in
accordance with the rotation curve based on maser sources,
see~\citealt{ras17}). Note that stars which were selected based on the
proximity of their parallaxes and proper-motion components lie along the
apparent main sequence in the colour--magnitude diagram (CMD)
(Fig.~\ref{fig:vdB130}, top). Thus this fact further corroborates the
existence of a stellar cluster in this region.

\begin{table*} \caption{The new list of the cluster vdB~130  members}
    \label{tab:new_vdB130}
    \begin{tiny}
        \begin{tabular}{clllllllllllr}
            \hline
            Source ID & RA (J2000)     &   DEC  (J2000) &     Plx &   e\_Plx & pmRA  &  e\_pmRA &  pmDE  & e\_pmDE &  G   & BP-RP & Dist & N \\
               &                &                &     mas &   mas   & mas/yr &  mas/yr   &  mas/yr & mas/yr   &  mag & mag   & kpc  &   \\
            \hline
            2061381510912036736 &20 17 45.3942 &39 22 50 &0.4611 &0.1158 &-3.380  &0.206 & -4.408 &0.191   &17.520  &2.412 &2.07&37\\
            2061377014100944000 &20 18 00.4084 &39 17 21 &0.4658 &0.1457 &-3.628 &0.229 & -4.465 &0.272   &18.078  &1.858 &2.06&38\\
            2061474591441143552 &20 18 13.7550 &39 24 57 &0.4660 &0.0937 &-3.426 &0.161 & -4.922 &0.192   &17.548  &2.051 &2.04&39\\
            2061471159769307008 &20 18 20.1225 &39 20 49 &0.4739 &0.0768 &-2.728 &0.131 & -4.533 &0.143   &17.245  &1.651 &2.00&40\\
            2061380484434531072 &20 18 02.3379 &39 20 30 &0.4766 &0.1087 &-4.098 &0.170 & -5.257 &0.183   &17.450  &2.081 &2.00&41\\
            2061474801901596928 &20 18 10.9328 &39 25 22 &0.4865 &0.0854 &-4.195 &0.143 & -5.833 &0.152   &17.310  &1.804 &1.96&42\\
            2061380686280218240 &20 17 52.6544 &39 21 09 &0.4919 &0.1657 &-3.806 &0.275 & -4.964 &0.276   &18.196  &2.253 &1.98&43\\
            2061380170883177472 &20 17 43.5947 &39 19 41 &0.4936 &0.1857 &-3.849 &0.299 & -4.419 &0.343   &18.324  &2.091 &1.99&44\\
            2061474355224990080 &20 18 09.9918 &39 22 23 &0.5014 &0.0762 &-3.369 &0.133 & -5.029 &0.143   &17.236  &1.948 &1.90&45\\
            2061380621873493376 &20 17 55.3558 &39 20 37 &0.5088 &0.0790 &-3.684 &0.127 & -5.422 &0.141   &16.946  &2.386 &1.87&a\\
            2061471430341784704 &20 18 15.8012 &39 22 27 &0.5095 &0.0444 &-3.293 &0.071 & -5.200 &0.081   &16.158  &1.912 &1.86&46\\
            2061381034190368256 &20 17 41.9580 &39 20 21 &0.5156 &0.0201 &-3.951 &0.030 & -4.747 &0.031   &13.204  &0.981 &1.84&13r\\
            2061474561383424000 &20 18 12.7681 &39 24 14 &0.5158 &0.1534 &-3.523 &0.275 & -5.025 &0.275   &18.056  &2.347 &1.89&47\\
            2061380415715060352 &20 17 55.9067 &39 20 07 &0.5176 &0.0942 &-4.163 &0.147 & -4.529 &0.153   &17.194  &2.645 &1.85&48\\
            2061380617558827520 &20 17 53.1780 &39 20 17 &0.5190 &0.1136 &-3.527 &0.179 & -5.334 &0.187   &17.597  &2.481 &1.86&49\\
            2061380003398199424 &20 17 49.7267 &39 18 07 &0.5208 &0.0439 &-3.887 &0.072 & -5.052 &0.077   &16.031  &1.944 &1.82&50\\
            2061380553146297728 &20 18 01.4642 &39 20 54 &0.5243 &0.1177 &-3.620 &0.189 & -5.081 &0.203   &17.772  &2.138 &1.84&51\\
            2061381102909847424 &20 17 42.8620 &39 21 11 &0.5256 &0.0331 &-3.794 &0.053 & -4.819 &0.053   &12.776  &1.099 &1.81&10r\\
            2061380381355314432 &20 18 01.6577 &39 20 02 &0.5258 &0.0324 &-3.900 &0.050 & -5.001 &0.053   &15.291  &1.389 &1.81&52\\
            2061381613991824256 &20 17 56.7649 &39 23 28 &0.5303 &0.6052 &-2.980 &1.030 & -5.007 &1.049   &18.443  &2.464 &2.16&53\\
            2061381343428026112 &20 17 39.2038 &39 22 38 &0.5307 &0.0194 &-3.803 &0.034 & -4.538 &0.031   &13.145  &0.990 &1.79&2r\\
            2061380037757937280 &20 17 53.7773 &39 19 13 &0.5312 &0.0532 &-3.458 &0.092 & -5.131 &0.097   &16.479  &1.862 &1.79&54\\
            2061381034190366208 &20 17 42.6727 &39 20 07 &0.5346 &0.0243 &-3.981 &0.050 & -5.212 &0.041   &14.494  &1.289 &1.78&14r\\
            2061471469006956032 &20 18 20.3443 &39 21 58 &0.5368 &0.0901 &-2.800 &0.153 & -4.895 &0.168   &17.325  &1.839 &1.79&55\\
            2061380656233230848 &20 17 56.1791 &39 20 46 &0.5392 &0.0237 &-3.857 &0.037 & -4.843 &0.041   &14.443  &1.664 &1.76&5r\\
            2061380514477932928 &20 18 06.4481 &39 21 17 &0.5395 &0.0269 &-3.562 &0.043 & -5.140 &0.047   &12.252  &0.845 &1.76&19\\
            2061380587513749888 &20 18 03.2004 &39 21 45 &0.5404 &0.0176 &-3.501 &0.028 & -5.208 &0.031   &13.013  &0.988 &1.76&18\\
            2061379964723777024 &20 17 55.7345 &39 18 18 &0.5409 &0.1486 &-3.794 &0.263 & -5.090 &0.293   &18.310  &2.380 &1.81&56\\
            2061380342680907136 &20 17 57.3517 &39 18 57 &0.5432 &0.0476 &-3.750 &0.078 & -5.356 &0.092   &16.255  &1.690 &1.75&57\\
            2061380312635856256 &20 17 47.0403 &39 20 19 &0.5447 &0.0204 &-3.700 &0.030 & -4.843 &0.036   &13.449  &1.009 &1.74&58\\
            2061381102909847296 &20 17 42.1893 &39 20 58 &0.5452 &0.0314 &-3.865 &0.047 & -4.96  &0.050   &10.052  &0.788 &1.74&11r\\
            2061474458304208512 &20 18 10.8129 &39 23 33 &0.5456 &0.0168 &-3.637 &0.028 & -5.231 &0.030   &13.523  &0.901 &1.74&17\\
            2061381068550105984 &20 17 44.6201 &39 21 05 &0.5473 &0.0538 &-3.693 &0.085 & -5.013 &0.109   &16.046  &2.022 &1.74&59\\
            2061381613991258880 &20 17 57.4966 &39 24 00 &0.5474 &0.1533 &-3.619 &0.253 & -4.596 &0.297   &18.319  &2.280 &1.79&60\\
            2061471400287485184 &20 18 12.9841 &39 21 36 &0.5479 &0.1228 &-3.043 &0.220 & -4.412 &0.248   &17.949  &1.772 &1.77&61\\
            2061380931111151360 &20 17 42.0249 &39 19 58 &0.5506 &0.0514 &-3.998 &0.082 & -4.824 &0.088   &15.935  &2.225 &1.73&62\\
            2061380450074795392 &20 18 00.5606 &39 20 38 &0.5523 &0.0256 &-3.437 &0.038 & -4.731 &0.045   &14.286  &1.465 &1.72&8r\\
            2061380243916371712 &20 17 52.8373 &39 20 02 &0.5533 &0.0233 &-4.097 &0.035 & -4.704 &0.036   &13.935  &1.337 &1.72&21\\
            2061379934678718976 &20 17 53.7659 &39 18 15 &0.5539 &0.0250 &-4.239 &0.041 & -5.268 &0.042   &14.627  &1.130 &1.72&6\\
            2061377529497025024 &20 18 05.2786 &39 19 56 &0.5604 &0.0290 &-3.931 &0.045 & -5.373 &0.050   &15.027  &1.235 &1.70&1\\
            2061474831956073088 &20 18 06.8633 &39 25 08 &0.5620 &0.0680 &-3.678 &0.118 & -5.212 &0.129   &17.031  &1.988 &1.70&63\\
            2061377254619104640 &20 18 11.7472 &39 18 48 &0.5621 &0.0402 &-3.909 &0.063 & -4.847 &0.071   &13.411  &0.940 &1.70&2\\
            2061380759312448384 &20 17 58.8207 &39 22 19 &0.5631 &0.0287 &-3.730 &0.044 & -4.961 &0.053   &12.567  &0.942 &1.69&9r\\
            2061381343428026496 &20 17 40.8205 &39 23 23 &0.5641 &0.0225 &-3.606 &0.036 & -5.180 &0.038   &14.620  &0.954 &1.69&64\\
            2061380553154011008 &20 18 02.0406 &39 21 18 &0.5655 &0.0911 &-3.900 &0.158 & -5.233 &0.169   &17.479  &2.250 &1.70&65\\
            2061380690592977280 &20 17 50.3958 &39 21 05 &0.5732 &0.0231 &-3.894 &0.035 & -4.772 &0.039   &14.234  &1.593 &1.66&1r\\
            2061376739215221888 &20 18  4.8007 &39 16 10 &0.5737 &0.0683 &-2.854 &0.114 & -5.477 &0.141   &16.705  &1.842 &1.67&66\\
            2061471258541187840 &20 18 21.4792 &39 21 21 &0.5738 &0.0237 &-3.654 &0.039 & -5.264 &0.042   &13.709  &0.986 &1.66&24\\
            2061377495137275520 &20 18 13.4588 &39 19 56 &0.5827 &0.0297 &-3.518 &0.046 & -4.966 &0.051   &15.078  &1.991 &1.64&67\\
            2061380858078916608 &20 17 59.0624 &39 23 24 &0.5857 &0.0334 &-3.726 &0.053 & -5.022 &0.058   &15.355  &1.873 &1.63&68\\
            2061380106477418880 &20 17 47.4214 &39 18 37 &0.5874 &0.0241 &-3.353 &0.038 & -5.456 &0.042   &14.453  &1.310 &1.62&4\\
            2061474733182122240 &20 18 04.8118 &39 24 38 &0.5886 &0.0220 &-3.780 &0.036 & -5.118 &0.039   &14.492  &1.825 &1.62&12\\
            2061380205240289152 &20 17 49.7836 &39 18 53 &0.5924 &0.0226 &-3.955 &0.035 & -5.222 &0.043   &13.000  &1.119 &1.61&4r\\
            2061377288978841472 &20 18 14.5332 &39 19 14 &0.5950 &0.0202 &-3.727 &0.030 & -5.344 &0.034   &13.706  &0.955 &1.60&69\\
            2061474423944472960 &20 18 07.0185 &39 23 18 &0.5985 &0.0230 &-3.596 &0.039 & -5.232 &0.042   &14.574  &0.993 &1.59&16\\
            2061380037757937152 &20 17 54.3807 &39 19 23 &0.6077 &0.0309 &-3.298 &0.048 & -5.274 &0.057   &15.179  &1.472 &1.57&20\\
            2061471121103942016 &20 18 16.4107 &39 19 42 &0.6143 &0.1642 &-3.003 &0.258 & -4.766 &0.296   &18.183  &1.927 &1.62&70\\
            2061380651920483712 &20 17 56.4076 &39 20 43 &0.6153 &0.1200 &-3.901 &0.199 & -5.158 &0.224   &17.711  &2.074 &1.59&5r\\
            2061380995516899584 &20 17 37.6178 &39 20 11 &0.6161 &0.1738 &-3.683 &0.311 & -4.995 &0.305   &18.464  &2.387 &1.62&71\\
            2061380518794267776 &20 18 07.0410 &39 21 41 &0.6175 &0.0556 &-4.027 &0.096 & -5.075 &0.105   &16.653  &2.049 &1.55&72\\
            2061378663368413440 &20 17 39.6536 &39 17 41 &0.6197 &0.0282 &-3.897 &0.044 & -4.966 &0.046   &14.759  &1.890 &1.54&73\\
            2061381751430215040 &20 17 44.0534 &39 24 13 &0.6274 &0.1520 &-2.946 &0.257 & -5.679 &0.290   &18.352  &1.999 &1.58&74 \\
            2061474698822384256 &20 18 01.4864 &39 23 55 &0.6461 &0.0300 &-3.691 &0.052 & -5.138 &0.053   &15.282  &1.702 &1.48&14\\
            2061381137269590016 &20 17 42.6726 &39 21 53 &0.6503 &0.0863 &-3.766 &0.136 & -5.149 &0.146   &17.142  &2.048 &1.49&75\\
            2061381137261972992 &20 17 43.6988 &39 21 47 &0.6604 &0.1344 &-4.274 &0.209 & -4.525 &0.216   &17.734  &2.189 &1.49&76\\
            2061380484434528384 &20 18 03.0200 &39 20 18 &0.6671 &0.0934 &-3.698 &0.145 & -5.220 &0.150   &10.819  &0.870 &1.46&3r\\
            2061381029874009088 &20 17 43.6459 &39 20 36 &0.7016 &0.0722 &-2.887 &0.111 & -4.517 &0.126   &10.923  &0.772 &1.38&12r\\
            2061380621873495424 &20 17 53.5368 &39 20 56 &0.6549 &0.0605 &-2.941 &0.098 & -5.793 &0.102   &15.008  &1.554 &1.53&22\\
            \hline
        \end{tabular}
    \end{tiny}
\end{table*}

These 68 stars can be considered to be probable members of vdB~130 cluster
inside a 12-arcmin sized area centred at the location mentioned above (see
Table~\ref{tab:new_vdB130}). These stars are identified using GAIA DR2 data
based on their positions on the CMD, closeness of their proper motions
(within 1 mas~yr$^{-1}$ of the mean value compared to 3 mas~yr$^{-1}$ when
identified using UCAC4 catalogue) and parallaxes (lying in the interval from
0.50 to 0.70 mas). Note that new population of the cluster contains 12 out of
14 stars (except 6r and 7r) initially identified~\citet{rac74} based
exclusively on analysis of their spectra and photometry. These stars are
marked by the letter `r'. Only 15 out of the remaining 36 stars from
table~2 in Paper~I met the parallax proximity selection criterion. The
missing numbers in the interval from 1 to 36 correspond to the UCAC4 stars
that failed to meet the selection criterion based on GAIA DR2 data. The
remaining stars are identified based on GAIA DR2 data exclusively and have
numbers ranging from 37 to 72. In Fig.\ref{fig:stars} are marked all stars
that meet all the selection criteria based on both catalogues. The 68 GAIA
DR2 stars mentioned above are most likely members of the rather compact young
cluster vdB~130 with a sky-plane diameter of about 6~pc.

In Table~\ref{tab:new_vdB130} we provide a new list of cluster members
extended mostly by including faint objects. Columns 1 gives source identifier
in GAIA DR2 catalogue; columns~2--3 give the J2000.0 equatorial coordinates;
columns 4--5, the parallaxes, uncorrected for systematical error, and appropriate
random errors; columns~6--9, the proper-motion components along both coordinates with their errors; column~10,
the $G$-band magnitudes; column~11, the $BP-RP$ colour indices, column~12,
the corresponding distances and the last column, the star numbers in the
cluster. Magnitudes $G$ and colour indices $BP-RP$ are rounded to reflect real photometric precision.

\begin{figure}
\includegraphics[width=\linewidth]{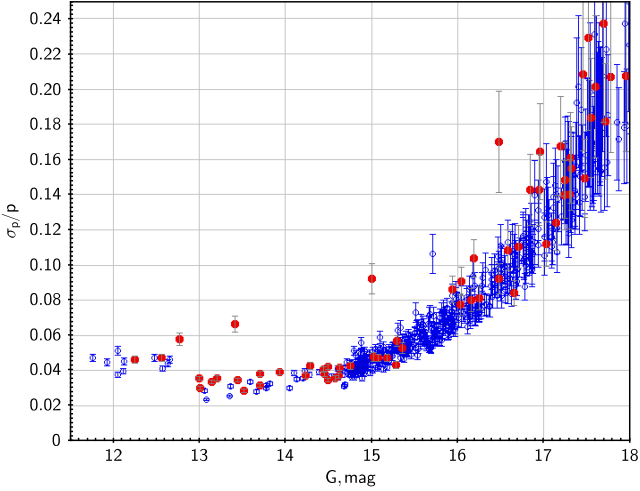}
\caption{Dependence of the relative error of GAIA DR2
trigonometric parallaxes on the G-band magnitude for members of
the vdB~130 (red) and NGC~188 (blue) open clusters.}
\label{fig:plx_rel_errors}
\end{figure}

Most of the stars with close proper motions (see
Fig.\ref{fig:vdB130}, bottom) lie in a rather broad line-of-sight
distance interval, from about 1.5 to 2.0~kpc. Such an elongated
shape of the cluster in the line-of-sight direction can be
explained entirely by random errors of GAIA DR2 trigonometric
parallaxes, which amount to  3--5 per cent even for the brightest
stars, translating into 50--80~pc in linear measure at the average
distance of about 1.7~kpc, and increase to 20--30 per cent for
faint stars (see Fig.\ref{fig:plx_rel_errors}, red symbols). Note
in this connection that  apparent radial extent due to parallax
errors is typical for many open clusters located at a distance
about 1.5--2~kpc and more when based on GAIA DR2 parallaxes. This
is immediately apparent from Fig.\ref{fig:plx_rel_errors} that
even for bona fide members of old open cluster NGC~188, selected
on the base of CMD, parallaxes and proper motions criteria (blue
symbols), located at a distance of about 1.5~kpc, relative errors
of GAIA DR2 parallaxes amount to 3--5 per cent for bright stars
and increase to $\sim$ 20 per cent for faint members of this
cluster. The relative error of trigonometric parallaxes evidently
increases with distance.

\subsection{The age of embedded cluster vdB 130 and the extinction law.}

Previously, we estimated the age of the vdB 130 cluster to be within
5-10~Myr (see Paper I) based on $BVJHK$ photometry and the theoretical
isochrone fitting on the CMD. However, it should be noted that determining
the age of young embedded clusters presents significant difficulties both
because of the differential extinction and the differences in the extinction
law on the way to different members of the cluster, associated with the
physical properties of the dust local to the cluster. Stars inside the
cluster are shifted on the CMD along the axes of colours and magnitudes
relative to their `true' positions. Moreover, for some stars this shift
occurs along the lines $\Delta M_V$ -- $\Delta (B-V)$ of different slopes and
lengths due to differences in the extinction law. For this reason, the
standard procedure for applying theoretical isochrone fitting technique to
the CMDs of such clusters cannot in principle give reliable results. Even if
there is a good estimate of the distance taken from the GAIA DR2
trigonometric parallax. Therefore, our age estimate published in Paper I can
be considered as a preliminary.

\begin{figure}
\includegraphics[width=\linewidth]{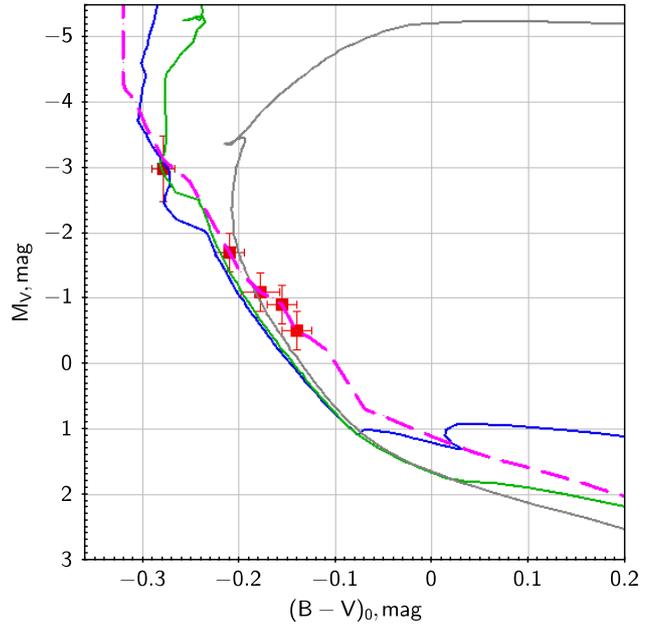}
\caption{ The colour $(B-V)_{0}$ -- absolute magnitude $M_{V}$ diagram for
hot stars in the vdB 130 cluster with a spectral classification. Blue, green,
and gray lines indicate Padova isochrons for ages 5, 10, and 30 Myr
\citep{Bressan}. Purple dashed line -- calibrations of luminosities and
colours of dwarf stars from Mamajek list (see text).} \label{fig:age}
\end{figure}

To estimate the age of the cluster vdB 130, we used the results of
spectral classification of 8 members of the cluster (1r, 3r, 4r, 5r, 11r,
12r, 19, 22), performed using data on the spectral energy distribution (SED)
from \citealt{tatar16}. All these stars belong to spectral classes B1, B1/B2,
B2/B3, B5, B5/B6 (in half of the cases neighboring spectral classes are
allowed). The normal colours and absolute values for stars of these spectral
classes were taken from~\citet{Pecaut} supplemented by Mamajek
list\footnote{\url {http://www.pas.rochester.edu/~emamajek/EEM_dwarf_UBVIJHK_colors_Teff.txt}}. Estimates of colour and magnitude errors were
made using original data on calibration stars presented by Mamajek
on the web-page\footnote{\url {http://www.pas.rochester.edu/~emamajek/spt/}}. The
positions of stars of spectral classes B1, B2, B3, B5, B6 are indicated on
diagram $M_{V}$ -- $(B-V)_{0}$ (see Fig.\ref{fig:age}).

The brightest star 11r has a spectral class $B1$. Based on its
position on CMD, the cluster is not older than $\sim$10 Myr. The large scatter
of fainter stars along the colour axis (see, for example, Figure 4 in Paper
I) even for IR data is explained not only by the differential absorption, but
also by the fact that the lifetime of low-mass objects at the protostar stage
can be comparable to the age of the cluster ($\sim$10 Myr). It can be clearly
seen on the Fig.\ref{fig:age} that in clusters with an age of 10~Myr, stars
of $A$ spectral class can still be on the pre-main sequence stage, and in the
clusters with an age of 5-10~Myr -- even some late $B$-stars. Thus, the age
of the embedded cluster vdB 130 is no more than 10~Myr. This estimate is
based on the SED analysis of eight $B$-stars in this cluster and is in
general agreement with our preliminary estimate given in Paper I.

Even if we know more or less reliable trigonometric distance derived
with the use of GAIA DR2 stellar parallaxes (true distance modulus $(m-M)_{0}
\approx 11.15$ mag), we can only determine the full absorption value for each
star, i.e. estimate differential absorption. In the absence of any data on
differential reddening, no age estimates are possible based on the position
of all cluster stars on the CMD. However, for the stars with known spectral
classes \citep{tatar16}, and with absolute magnitude and colour calibrations
by Mamajek mentioned above, we have attempted define the ratio $R_{V} = A_{V}
/ E (B-V)$ , which characterizes the extinction law. In the
Table~\ref{tab:ages} the absorption values of $A_{V}$ and $A_{K}$, as well as
estimates of $R_{V}$ and $A_{K} / E(J-K)$ ratios for stars 1r, 3r, 4r, 5r,
11r, 12r, 19, 22 are given. The photometric data in the $BVJHK$ bands are
taken from the table 2 in Paper I. 

\begin{table*}
\caption{Members of vdB~130 cluster with spectral classification}
\label{tab:ages}
\begin{scriptsize}
\begin{tabular}{l|l|l|l|l|l|l|l|l}
\hline
  & Sp & $E(B-V)$(T)& $E(B-V)$(PM)& $A_{V}$& $R_{V}$& $E(J-K)$&  $A_{K}$& $A_{K}/E(J-K)$\\
  &    & mag        & mag         & mag    & mag    & mag     & mag     & mag       \\
\hline

 1r&  B1V&      1.3& 1.14&       6.06&    5.3&       1.40&        1.48&         1.06\\
 3r&  B1V&      0.9& 0.86&       2.85&    3.3&       0.47&        0.51&         1.06\\
 4r&  B1V/B2V&  1.1& 1.01/0.94&  5.1/3.7& 5.0/3.9&   0.68/0.62&   2.0/0.9&      2.9/1.4\\
 5r&  B1V/B2V&  1.4& 1.28/1.21&  6.7/5.4& 5.2/4.5&   1.09/1.03&   2.2/1.1&      2.0/1.0\\
11r&  B1V&      1.1& 0.82&       2.04&    2.5&       0.44&      $\approx0$&     $\approx0$\\
12r&  B5V&      0.5& 0.68&       0.80&    1.2&       0.36&       $\approx0$&   $\approx0$\\
19&   B2V/B3V&  0.9& 0.79/0.76&  3.0/2.4& 3.8/3.1&   0.40/0.37&   0.8/0.3&     2.0/0.8\\
22&   B5V/B6V&  1.4& 1.25/1.23&  5.0/4.6& 4.0/3.7&   0.92&        1.3/0.9&     1.4/1.0\\
\hline
\end{tabular}
\end{scriptsize}
\end{table*}

Columns 2 and 3 give the spectral class and the colour excess
estimated by \citealt{tatar16} using data on the spectral energy distribution
(SED). Columns 4 and 7 show colour excess values $E(B-V)$ and $E(J-K)$,
defined as the differences between the observed colours and colours taken
from colour calibrations \citep{Pecaut} for the spectral classes specified in
column 2. The absorption values in the bands $V$ (column 5) and $K$ (column
8) were estimated using obvious expressions $A_V$ = $(V-M_V)-~(m-M)_{0}$ and
$A_K$ = $(K-M_K)-~(m-M)_{0}$, where true distance modulus $(m-M)_{0} \approx
11.15$ mag. Finally, columns 6 and 9 give the ratios $R_V = A_{V} / E(B-V)$
and $R_K = A_K / E(J-K)$. From the uncertainties of colour calibrations and
true distance modulus, their errors can be estimated as $\approx 0.2$. For
stars with two variants of spectral classification, both values are given.
Note that the colour excess values $E(B-V)$ defined by two different ways
agree fairly well. It is noteworthy to note that the ratio $R_V$ for stars
1r, 4r, 5r, 22 is significantly higher than the `standard' value $R_{V}
\approx 3.1$ for the `conventional' extinction law \citep{Cardelli}. For
stars 11r, 12r, the absorption $A_{K}$ within the errors is nearly zero,
which is confirmed by the relatively small absorption $A_{V}$ and the
unusually small value of $R_V$ for star 12r. For stars 3r, 11r, 19, the ratio
$R_{V}$ can be considered as normal within the errors. The data shown in the
Table~\ref{tab:ages} directly confirm the presence of abnormal absorption in
the vdB 130 cluster area (see also \citep{tatar16}). It should be noted that
for most stars, with the exception of 11r, 12r, the ratio $A_{K} / A_{V}$ is
approximately twice the `normal' value $\approx 0.1$ ~\citep{Cardelli}.

\subsection{Protocluster region.}

\begin{figure}
\includegraphics[width=0.8\linewidth]{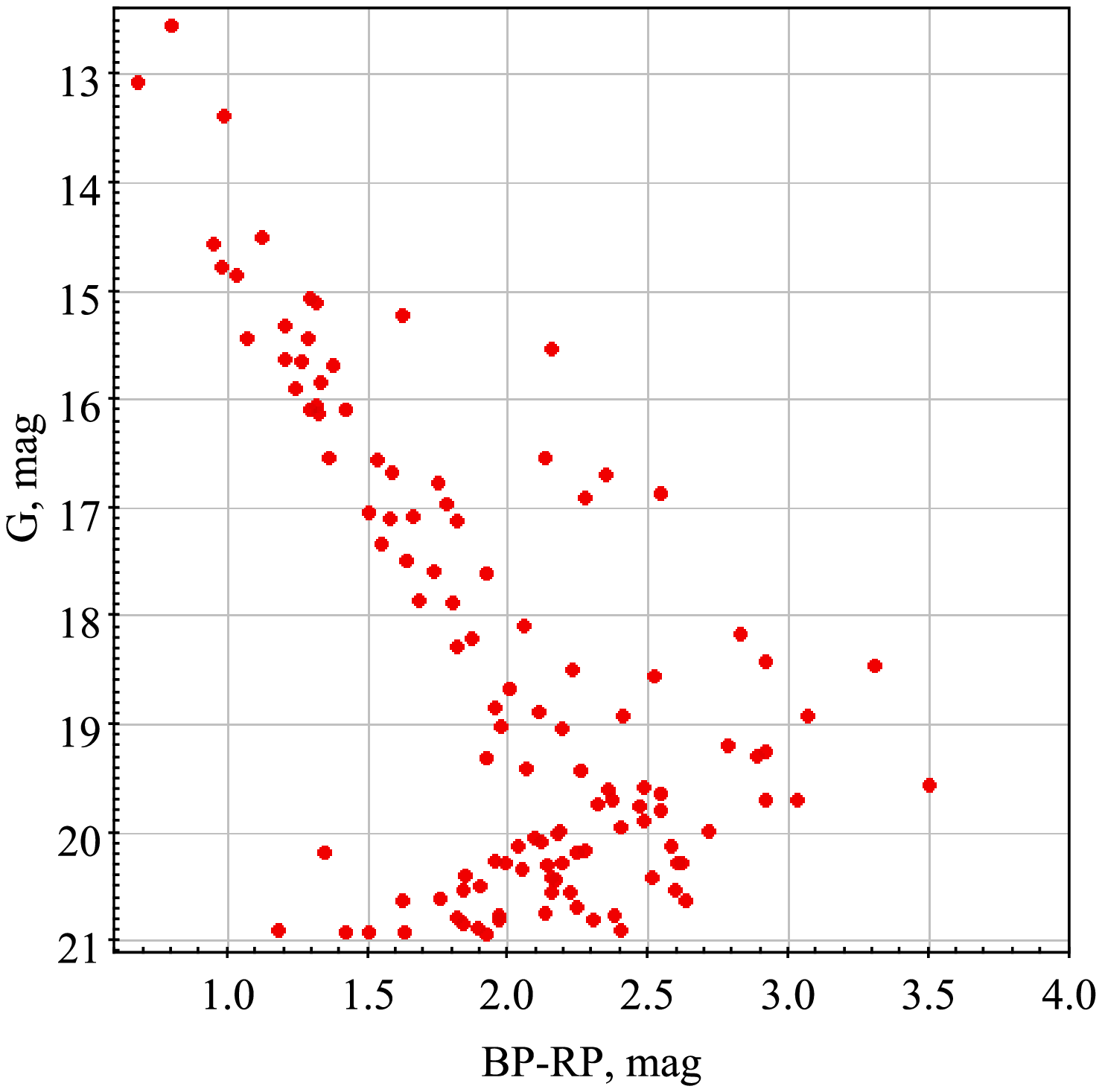}

\includegraphics[width=0.8\linewidth]{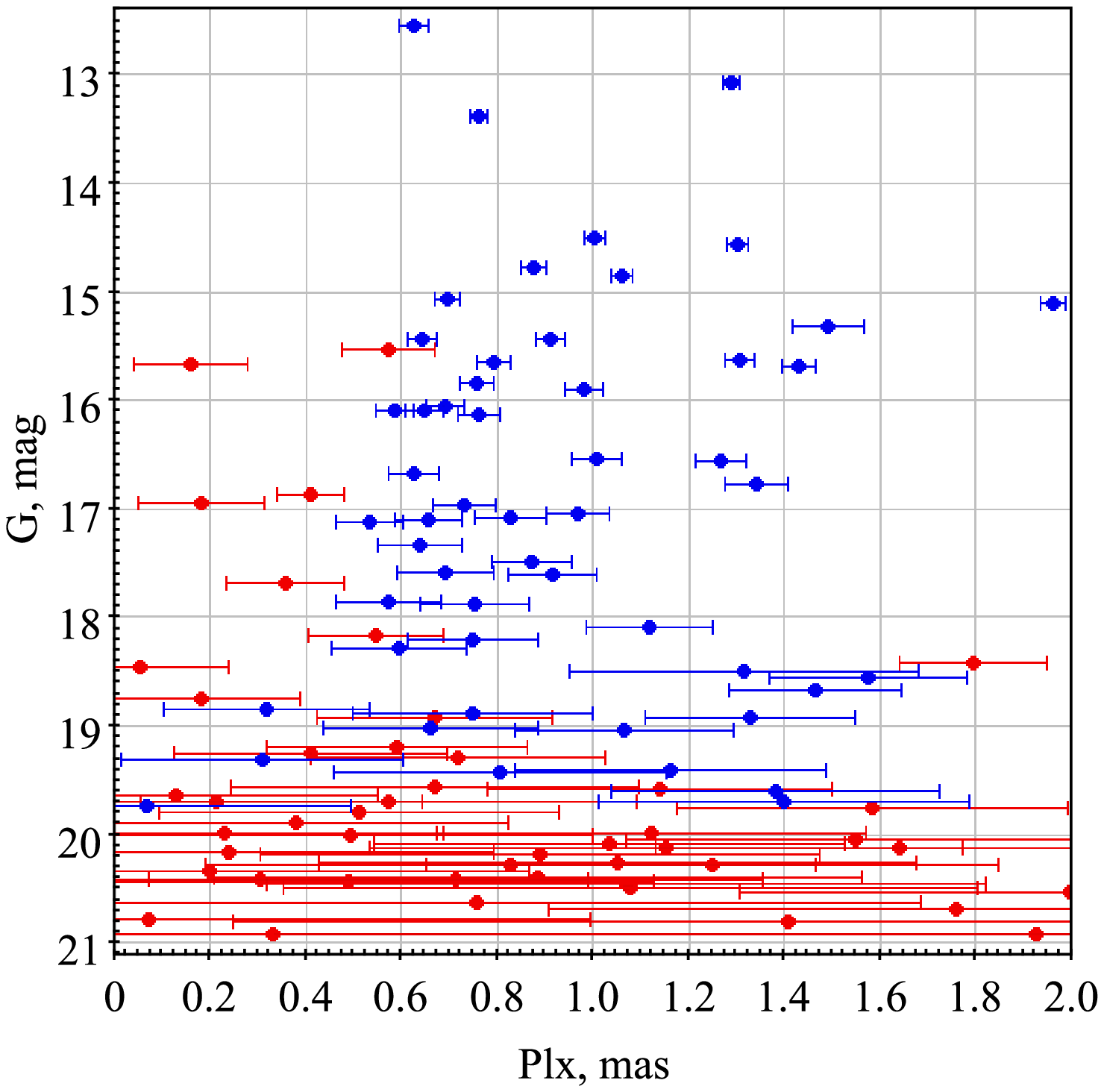}

\includegraphics[width=0.8\linewidth]{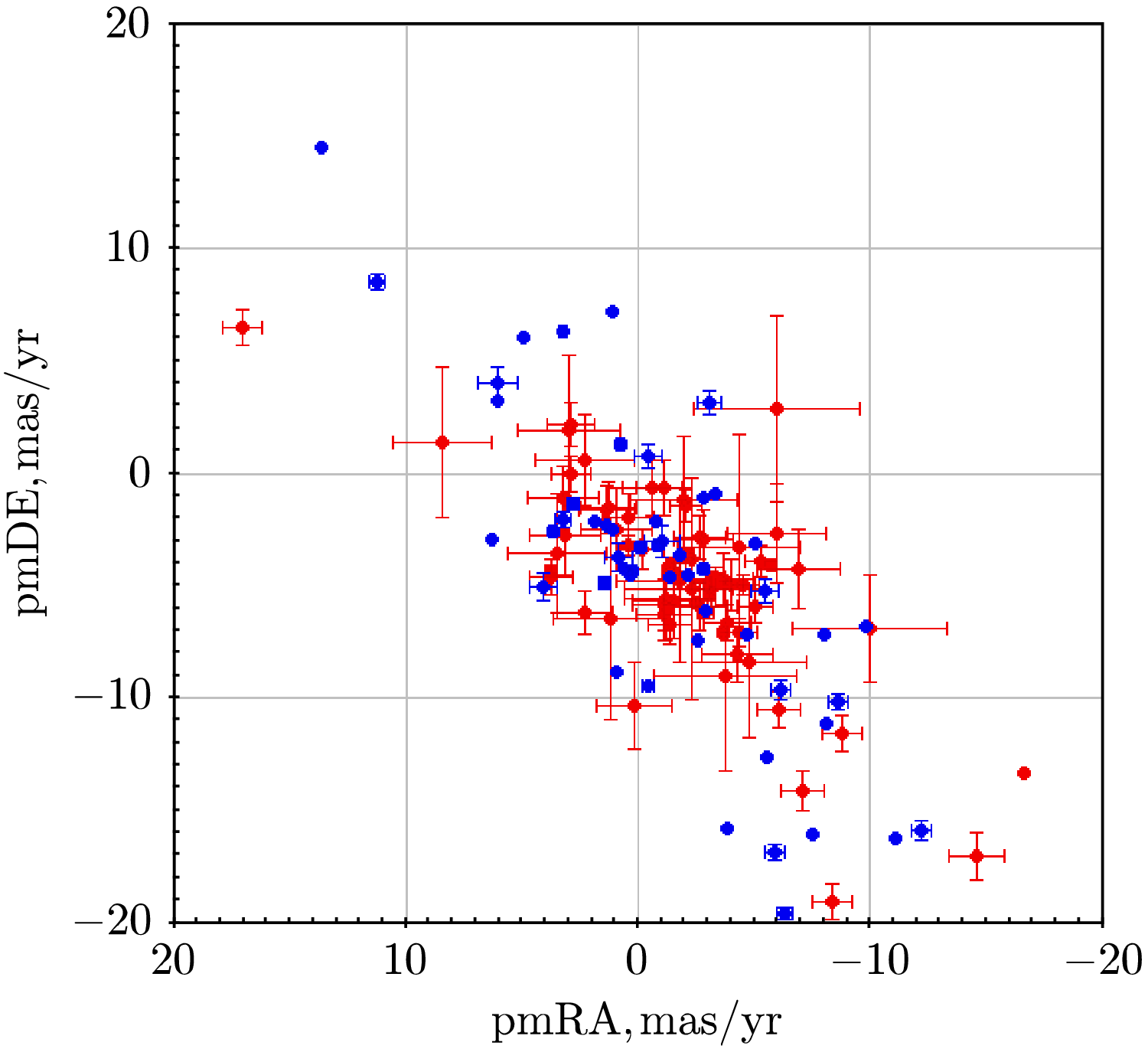}
\caption{ The same as Fig. 3 but for 137 stars in the protocluster area.
The $(BP-RP)$ -- $G$ colour-magnitude diagram (top panel),
parallax--magnitude diagram (middle panel) and two-dimensional proper-motion
diagram (bottom panel) for stars in the direction of the protocluster.
Blue symbols (middle and bottom panel) indicate the stars that form the
apparent Main Sequence.} \label{fig:proto}
\end{figure}

In the direction toward a dense condensation in  the tail of the molecular
cloud a compact group of class I and II protostars was found
(Fig.~\ref{fig:cloud-proto}, bottom, see also Paper II). While searching for
optical counterparts of the protocluster we investigated the 2.5-arcmin
radius area centred on the protocluster (
$\alpha_{2000}\sim20^{\mathrm{h}}17^{\mathrm{m}}00^{\mathrm{s}}$,
$\delta_{2000}\sim 39^{\circ} 21^{\prime}00^{\prime\prime}$) and found 137
GAIA DR2 stars. Like in the case of vdB~130 cluster we plotted the
Hertzsprung-Russell (Fig.~\ref{fig:proto}, top), parallax -- $G$-band
magnitude (Fig.~\ref{fig:proto}, middle) and the two-dimensional
proper-motion diagrams (Fig.~\ref{fig:proto}, bottom) for stars in the
direction of the protocluster. As  follows from an analysis of the figures
the optically observed stars  making  up the apparent main sequence (Fig.
\ref{fig:proto},top) span a broad interval of heliocentric distances ranging
from 0.7 to 2~kpc (Fig.~\ref{fig:proto}, middle). Moreover, these stars also
show no appreciable concentration in the proper-motions diagram
(Fig.~\ref{fig:proto}, bottom) and, in our opinion, do not form a
gravitationally bound group of objects.

In Paper I we proposed the following scenario of the ongoing
evolution of this region  (see also \citep{smith}). In the cluster
area, the expanding supershell around Cyg~OB1 interacts with the molecular
cloud. A typical cometary shape of the cloud (an IR pillar), (see
Fig.~\ref{fig:cloud-proto}), is an indirect confirmation of such interaction.
The wind and the UV radiation of the Cyg OB1 stars had triggered the star
formation in the pillar, having resulted in the emergence of the vdB~130
cluster. The compact protocluster observed near a dense clump in the tail of
the cometary molecular cloud is the next burst of star formation.

\section{Spectral types and colour excesses of the stars studied.}

Although we could not identify a condensation of objects with
close proper motions in the protocluster region, we selected 25 stars for
spectroscopic study. Their proper motions do not strongly
differ from those of the stars of the vdB~130 cluster located in
the same molecular cloud. Note that the stars studied are located
within 2.5~arcmin from the protocluster centre and are projected
onto the break in the star-formation filament observed there.
This filament can be traced in the cometary cloud by an ensemble
of several millimetre-wave sources ~\citep{motte}. It is also
described in Paper II (see fig.13 there) based on \textit{Herschel} space telescope
data.

Table~\ref{tab:spectr_class} presents the results of our analysis of
spectroscopic and photometric data for 25 stars. Column~1 gives the
designation of the star according to Fig.~\ref{fig:cloud-proto1}; column~2,
the spectral type determined in this study; column~3, the observed $(B-V)$
colour index; column~4, the intrinsic colour index characteristic of this
spectral type~\citep{str}; column~5, the $E(B-V)$ colour excess; column~6,
the $(J-K)_{calc}$ colour index computed taking into account the inferred
spectral type and colour excess assuming normal interstellar extinction law;
column~7, the observed colour index $(J-K)_{obs}$; column~8, the absolute
magnitude $(M_V)_{calc}$ determined using the distance to the object (see
column~8 in Table~2) and the colour excess $E(B-V)$ assuming normal
interstellar extinction law, and column~9 gives the average absolute
magnitude $M_V$ for the spectral types listed in column~2~\citep{gray}.

\begin{table*} \caption{Physical properties of stars observed toward the protocluster.
Column~1~--- star numbers; column~2~--- spectral types; column~3~--- observed
$(B-V)$ colour indexes; column~4~--- intrinsic colour indexes; column~5~---
$E(B-V)$ colour excesses; column~6~--- calculated $(J-K)$ colour indexes;
column~7~--- observed $(J-K)$ colour indexes; column~8~--- calculated
absolute magnitudes $M_V$; and column~9~--- average absolute magnitudes $M_V$
for the spectral types listed in column~2 } \label{tab:spectr_class}
\begin{scriptsize}
\begin{tabular}{l|l|l|l|l|l|l|l|l}
\hline
Star & Sp &$(B-V)_{obs}$ &$(B-V)_{0}$ &E(B-V) &$(J-K)_{calc}$ & $(J-K)_{obs}$ & $(M_V)_{calc}$  & $M_V$ \\
     &    & mag          & mag        & mag   & mag           & mag           & mag             & mag \\
\hline
 2  &  A5 - A7V  &0.73& 0.15 - 0.20& 0.58 - 0.53     & 0.39 &       0.36 & 1.1 - 1.3  & 1.2(A5IV)\\
 3  &  F3V      &0.51&     0.40    &     0.11        & 0.30 &       0.26 &    3.3      & 3.1(F3V)\\
 5  &  G2 - G5V &0.79& 0.62 - 0.68 & 0.17 - 0.11     & 0.45 &       0.45 &  4.6 - 4.8   & 4.7(G2V)\\
 6  &  K4V      &0.99&     1.05    & $\simeq0$       & 0.66 &       0.72 &     6.9      & 7.1(K4V)\\
 7  &  G6 - G8V &0.71& 0.70 - 0.75 & $\simeq0$       & 0.42 &       0.42 &      5.3     & 5.3(G6V)\\
 8  &  G2 - G5V &0.91& 0.62 - 0.68 & 0.30 - 0.24     & 0.52 &       0.57 &  4.5 - 4.6 & 4.7(G2V)\\
10  &  F6 - F7V &1.04& 0.46 - 0.50 & 0.55 - 0.59     & 0.54 - 0.61& 0.56 &      3.5     & 3.7(F6V)\\
11  &  F2 - F3V &1.32&     0.36    & $\simeq1.0$     & 0.77&       1.55  &    2.0       & 1.9(F3IV)\\
12  &  A5 - F3V &0.91& 0.30 (F0V)  &     0.6         & 0.49 &       0.49 &     2.8      & 2.6(F0V)\\
13  &  G2 - G5  &1.06& 0.62 - 0.68 & 0.4 - 0.34      & 0.56 - 0.51& 0.52 & 4.6 - 4.7  & 4.79(G2V)\\
14  &  F3V?     & 1.56&     0.40   & $\simeq1.2$     & 0.87 &      1.21  &     2.2      & 1.9(F3IV)\\
15  &  M0 - M1V &1.39& 1.44 - 1.47 & $\simeq0$       & 0.89 &      0.89  &     8.9      & 9.2(M0V)\\
17  &  K2 - K4III&1.75& 1.16 (K2III)&     0.6        & 1.04 &      1.09 &     0.5     & 0.6(K2III)\\
18  &  G6 - G8V &1.12&     0.75    & 0.37            & 0.62  &      0.62 &     5.1     & 5.3(G6V)\\
19  &  M1 - M2V &1.37& 1.47 - 1.49 & $\simeq0$       & 0.9   &      0.87 &     9.5     & 9.7(M1V)\\
20  &  K1 - K2V &0.90& 0.86 - 0.91 & $\simeq0$       & 0.54 - 0.59& 0.61 &     6.4     & 6.3(K2V)\\
21  &  G2 - G5V &1.22& 0.62 - 0.68 & 0.66 - 0.6      & 0.71  &      0.63 &  4.3 - 4.5 & 4.7(G2V)\\
22  &  G5 - G8V &0.97&0.68 - 0.75 & 0.3  - 0.23      &0.54   &      0.47 &   3.6 - 3.8  & 3.2(G5VI)\\
23  &  F3 - F5V &0.73&0.40 - 0,44 & 0.33 - 0.29    &0.35 - 0.26& 0.35  &   3.7 - 3.8 & 3.1(F3V)\\
\hline
\end{tabular}
\end{scriptsize}
\end{table*}

\begin{figure}
\includegraphics[width=\linewidth]{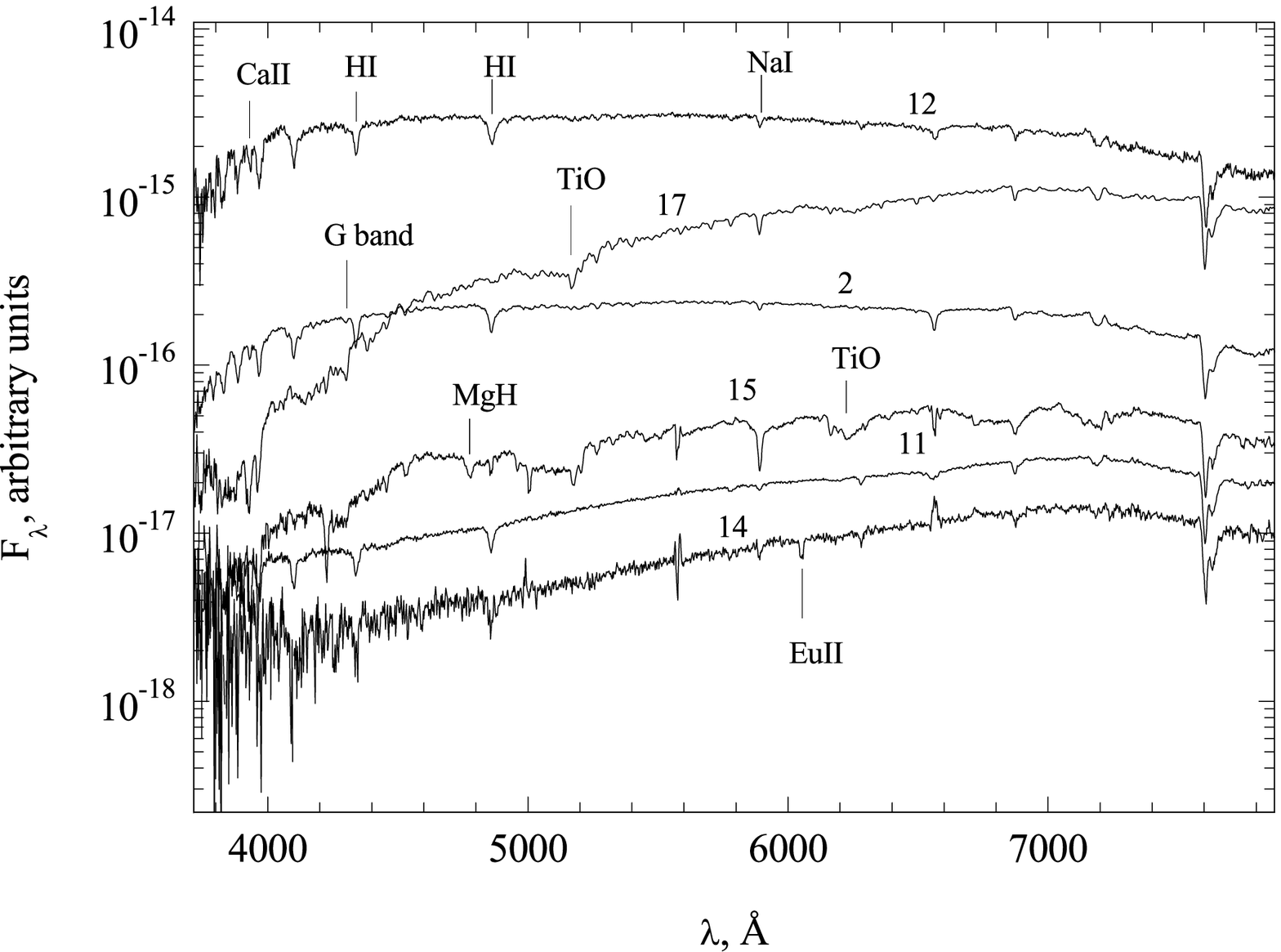}
\caption{Examples of stellar spectra. The numbers of stars from
Table~\ref{tab:spectr_class} and spectral lines used for determining spectral
classes are indicated.} \label{fig:fig_sps}
\end{figure}

As is evident from Table~\ref{tab:spectr_class} (see columns~6 and~7), we
achieved good agreement between the computed and observed colour indices
$(J-K)$ for all the stars studied except two objects -- stars~11 and~14
(Fig.~\ref{fig:fig_sps}). Such a significant discrepancy may be due to
circumstellar envelopes surrounding these stars. Indications of circumstellar
matter surrounding star 11 can also be seen in direct  $K$-band images. Note
that the distance to star 11 agrees, within the errors, with the distance to
the molecular cloud, whereas according to GAIA DR2 data, star 14 must be
located behind the cloud (see column~9 in Table~\ref{tab:stars}). These stars
exhibit certain spectral peculiarities. The DIBs can be seen in the spectrum
of star 11, whereas the spectrum of star 14 shows a Eu~\textsc{ii} line
(possibly blended with a  Nd~\textsc{ii} line, see Fig.~\ref{fig:fig_sps}).

The star 2 (an A5-type subgiant) has the earliest spectral type among the
observed stars (Fig.~\ref{fig:fig_sps}). Spectral classification of A-type
stars is usually based on an analysis of Balmer lines. In our case it is very
difficult to perform because the bright and highly non-uniform nebula also
shines in hydrogen lines. As a result of the background subtraction, hydrogen
absorption lines present in the spectrum of an A-type star may appear both
stronger and weaker than in reality. However, there is another criterion for
classification of A-type stars -- it uses Ca~\textsc{ii} H and K lines. The fact that
this is not a B-type star is evident from the absence of He~\textsc{i} lines in its
spectrum. The lack of the G-band in the spectrum and rather weak Ca~\textsc{i} 4226~\AA\,
line indicate that we are not dealing with an F-type star.

Most of the stars (13) are of spectral types G and F. We classified G-type
stars based on the depth of the G-band and intensity of the Mg~\textsc{i} 5167, 5172,
and 5183~\AA\, lines (Fig.~\ref{fig:fig_sps}). There are a total of seven such
stars and the remaining six stars are of spectral type F.

The remaining stars are of spectral type $K_{s}$ and early M (see
Table~\ref{tab:spectr_class}). We classified them based on the properties of
MgH- and TiO-bands and the Na~\textsc{i} 5890, 5896~\AA\, doublet
(Fig.~\ref{fig:fig_sps}). The latter, however, may be somewhat distorted
because of superimposed interstellar Na~\textsc{i} lines, but almost all such
stars have small colour excesses (see column~3 in
Table~\ref{tab:spectr_class}), and this effect must be weak. Note that K- and
M-type stars can be rather accurately subdivided into luminosity classes V
and III.

At the last stage we compared our inferred $V$-band absolute magnitude with
the $M_V$ estimate averaged over  the spectral type and luminosity class
presented by~\citet{gray}. Note that we estimated the absolute  $V$-band
magnitude based on the photometric measurement, distance, and interstellar
extinction. For several object the best agreement is achieved by assuming
that the star is of luminosity class IV rather than  V as we initially
believed. One must, however, bear in mind that the scatter of absolute
magnitudes of different stars within the same spectral type and luminosity
class may amount to 1~mag, which is comparable to the difference between the
average  $M_V$ values for stars of luminosity class IV and V. For example,
the corresponding difference for A5--A9 type stars is of about 0.8--0.9 mag.

\begin{figure}
\includegraphics[width=\linewidth]{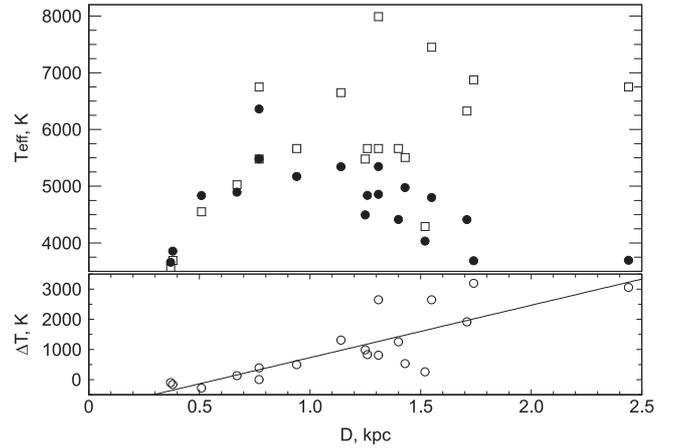}
\caption{Distance dependence of the effective temperature derived
from spectral type (the squares) and adopted from GAIA DR2 catalogue (the
filled circles)~--- top panel, and distance dependence of the difference
between these temperatures (the unfilled circles)~--- bottom panel.}
\label{fig:temp-dist}
\end{figure}

The data from Table~\ref{tab:spectr_class} can be used to compare the
effective temperatures of stars in the protocluster region inferred from
spectral types \citep{gray} with the temperatures provided in the GAIA DR2
catalogue. Fig.~\ref{fig:temp-dist} shows the dependence of pointed above
quantities from heliocentric distances to stars. It is seen from the figure
that the temperatures derived using different methods agree well up to a
distance of $\sim 0.8$ kpc. Note that the effective temperatures determined
from spectra in this paper may be both lower and higher than GAIA DR2
estimates. Note also that according to our observational data the
interstellar extinction toward the stars mentioned above located within
$\sim0.8$~kpc is close to zero (see the text below and
Table~\ref{tab:spectr_class}). The difference increases with the distance
(and hence with interstellar extinction) -- the values of temperature
provided by GAIA DR2 becomes systematically lower than our estimates.
Consider, for example, stars 11 and 14, which are of spectral type F
according to our estimates (Table~\ref{tab:spectr_class}). The GAIA DR2
estimates of their temperatures (about 3700~K) correspond to early  M type
and hence their spectra should be dominated by molecular bands. However, as
is evident from Fig.~\ref{fig:fig_sps}, this is not  the case. This fact is
most likely indicative of insufficient account of the effect of interstellar
extinction in GAIA DR2.

\section{Interstellar extinction toward the molecular cloud.}

\begin{figure*}
\includegraphics[width=0.49\linewidth]{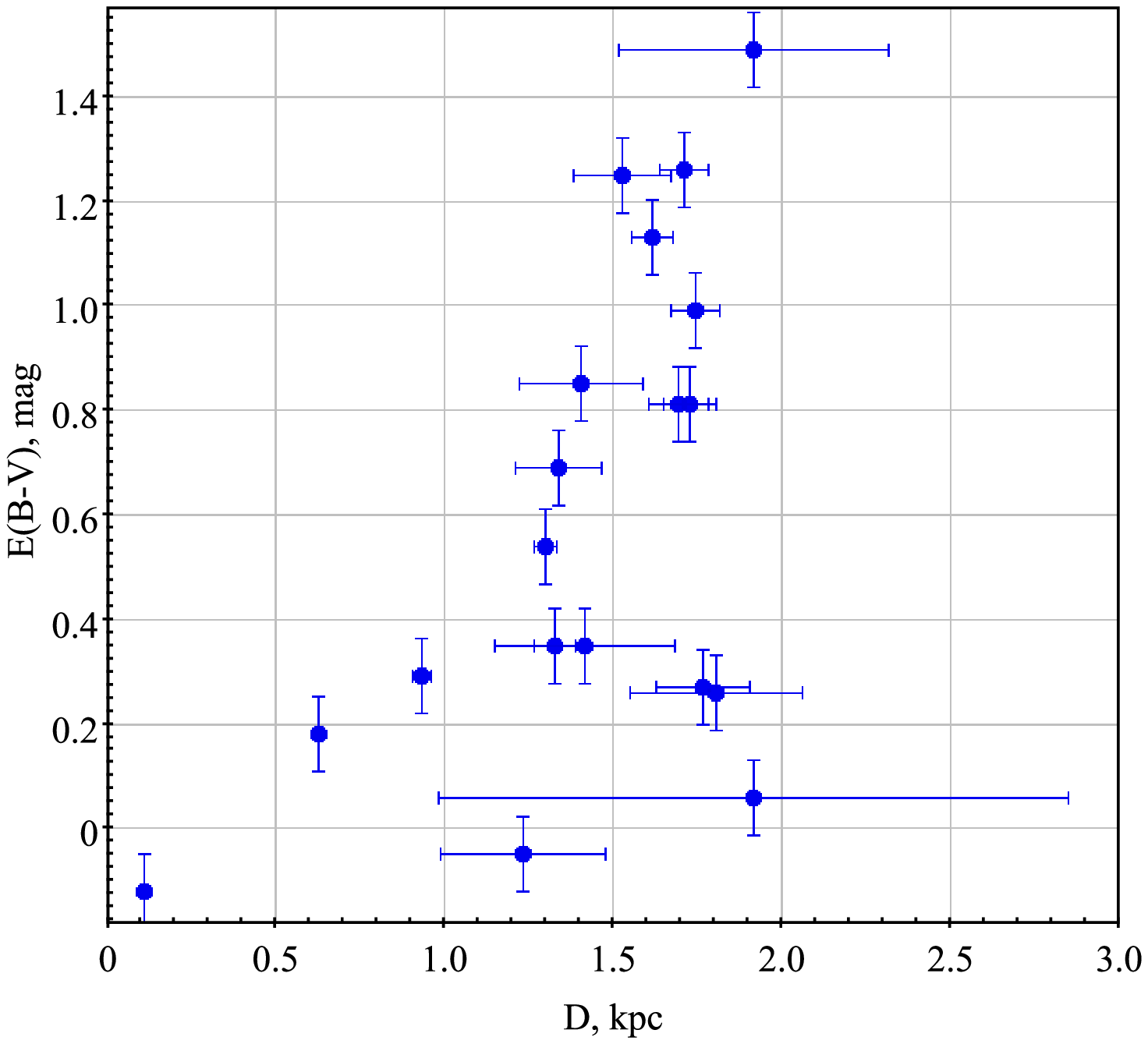}
\includegraphics[width=0.49\linewidth]{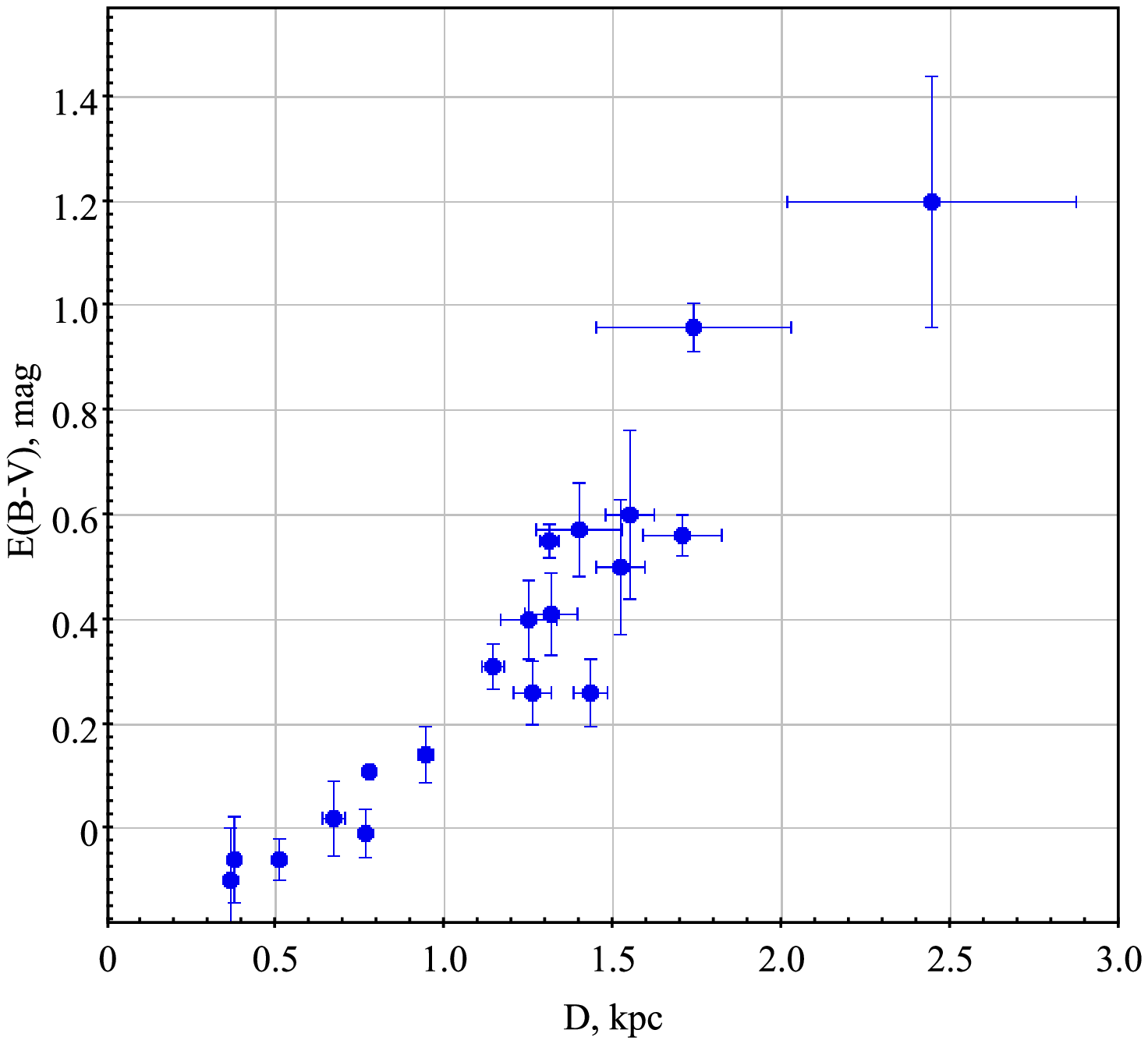}
\caption{Dependence of the colour excess $E(B-V)$ on the distance $D$ toward vdB 130
cluster (left) and the protocluster (right). Distances are based on
GAIA DR2 parallaxes corrected for systematic error. Left panel shows not only
some vdB~130 cluster members but also field stars.} \label{fig:ebv}
\end{figure*}

\begin{figure*}
\includegraphics[width=0.49\linewidth]{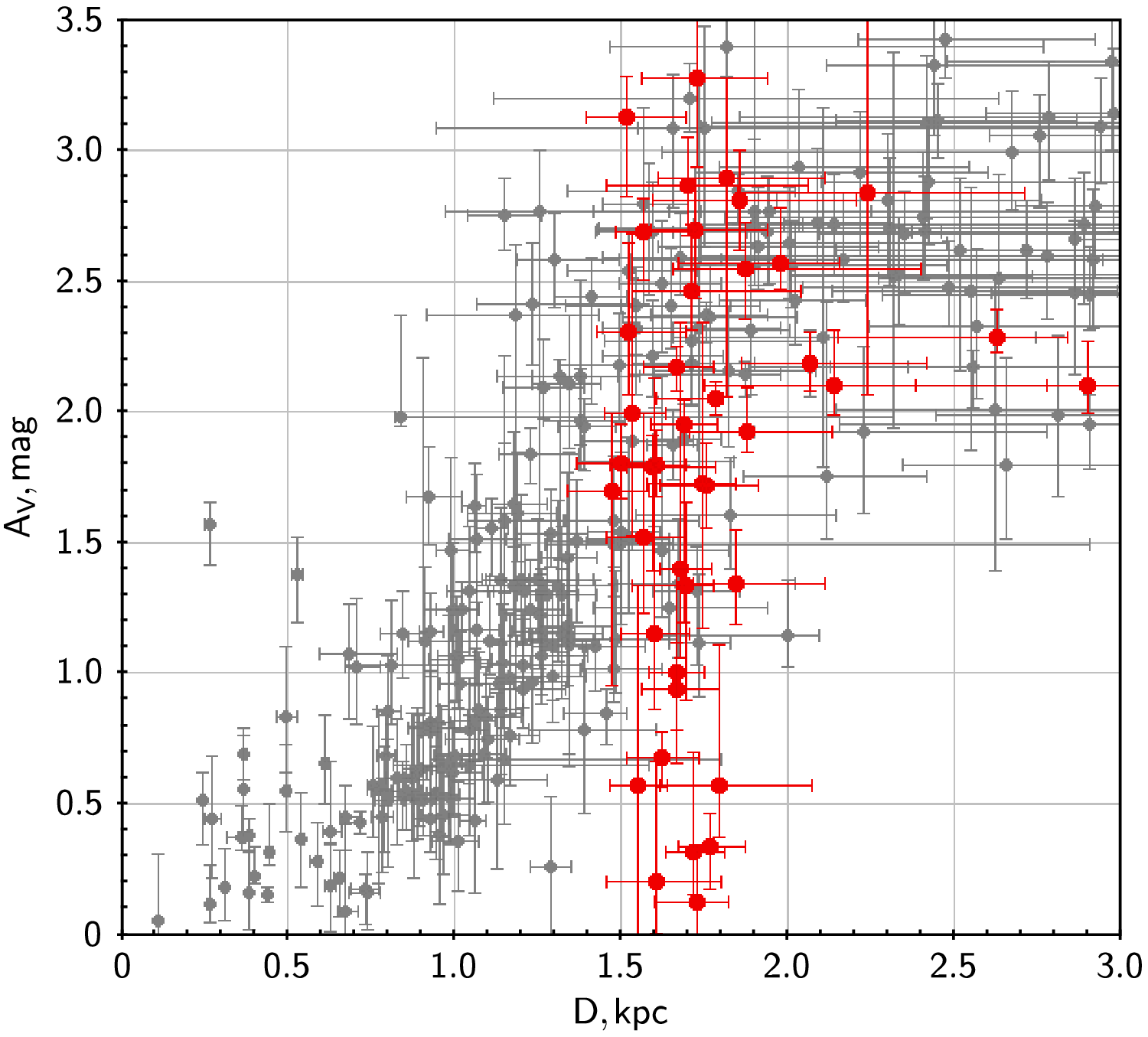}
\includegraphics[width=0.49\linewidth]{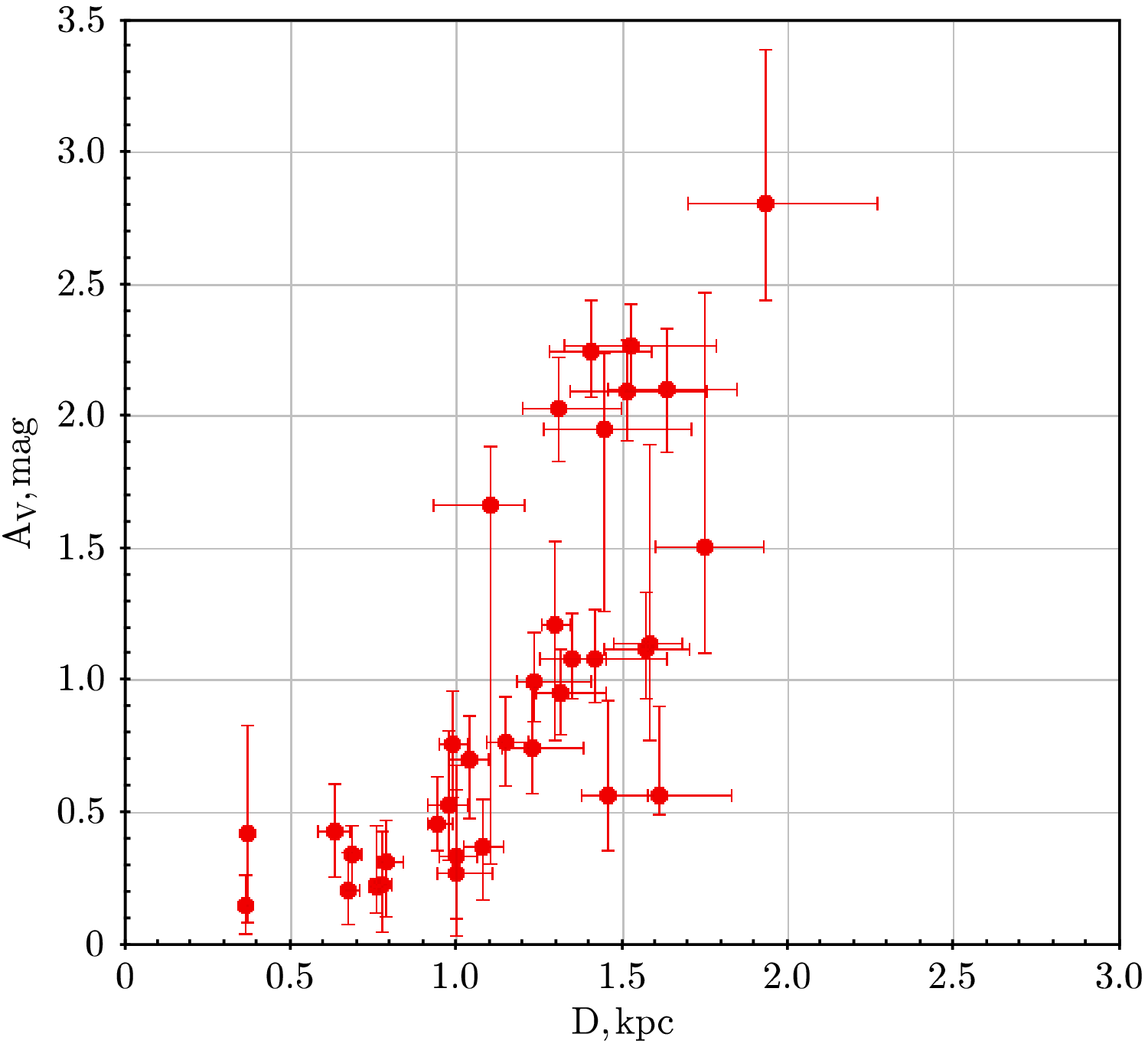}
\caption{The same as on Fig.~\ref{fig:ebv} but for the extinction $A_V$ and distance $D$
taken from the StarHorse catalogue \citep{anders}. Left panel: vdB~130 field. Cluster members
listed in the Table~\ref{tab:new_vdB130} are marked by red, and field star are marked by grey.
Right panel: protocluster field. See Section 5 for details.} \label{fig:Av}
\end{figure*}

Right panel of Fig.~\ref{fig:ebv} shows the dependence of $E(B-V)$ on
the heliocentric distance to stars toward the protocluster
based on the data from Table~\ref{tab:stars} and Table~\ref{tab:spectr_class}.
The lack of appreciable reddening toward
the protocluster out to the distances of 0.6--0.8~kpc is immediately
apparent. This agrees with the data from~\citet{Neckel} for this region
and explains the agreement between GAIA DR2 effective temperatures and
our temperature estimates based on spectral types within the distance limit
mentioned above. In the case of normal interstellar extinction law ($R_V
\approx 3.1$) extinction at larger distances increases at a rate of about
1.9 mag~kpc$^{-1}$, which is in general agreement with average rate in
the disk of the Milky Way galaxy \citep{Sharov}. Note that star 11 is the only one to deviate
significantly from this dependence. The observed  $(J-K)$  colour index is
much redder than calculated on the base of $E(B-V)$ colour excess under the assumption of
normal extinction law (see Table~\ref{tab:spectr_class}). This object
probably has a circumstellar dust shell distorting the energy distribution in
the spectrum of the star. This hypothesis is supported by the fact that star
11 is a bright source on 3.6--8~$\mu$m images taken by \textit{Spitzer} space
telescope. According to Table~\ref{tab:stars}, the observed colour index
$(J-K)$ of star 14 is also inconsistent with the colour index computed from
$E(B-V)$ based on standard extinction law. However, the  $E(B-V)$ is
consistent with the distance to the star (see Fig.~\ref{fig:ebv}, right).
This discrepancy may be indicative, e.g., of the presence of a circumstellar
disk seen face on.

For comparison, we show in Fig.~\ref{fig:ebv} (left)  the dependence of
$E(B-V)$ on distance for stars of the open cluster vdB~130 and field stars
(with the parallaxes corrected for the systematic error). To estimate
reddening values we used the spectral types and observed $(B-V)$ colour
indices from~\citep{tatar16} and adopted intrinsic colour indices
from~\citep{str}. Three stars are immediately apparent, which are located at
the cluster distance of $\sim1.8$~kpc and have anomalously low reddening
$E(B-V)<0.3$. According to GAIA DR2 parallaxes and proper motions, two of
these stars `a' and 50 are members of the vdB 130 cluster (see
Table~\ref{tab:new_vdB130}). Their small $E(B-V)$ colour excess may be due to
the fact that these are red dwarfs \citep{tatar16} with magnitudes $B=19 -
22$ mag, for which the possible $E(B-V)$ errors may amount to several tenths
of a magnitude.

Like in the case of the protocluster, we can see on Fig.~\ref{fig:ebv}
(right panel) the lack of appreciable reddening out to a distance of about
$\approx 1$ kpc. For more distant stars reddening increases faster with
distance than we see in Fig.~\ref{fig:ebv} (right). This fact is indicative
of extra extinction in the region of the young open cluster vdB~130.

By combining GAIA DR2 ($G$, $BP$, $RP$), PanSTARRS1, 2MASS and
AllWISE photometry, \cite{anders} calculated new $A_V$ extinction values and
made new Bayesian estimates of photo-astrometric distances for approximately
265 million stars brighter than $G = 18$~mag. We tried to qualitatively
compare these new data taken from StarHorse catalogue by \cite{anders} with
our data on colour excess and its variation along line-of-sight.
Fig.~\ref{fig:Av} shows the dependence of the extinction $A_V$ on the
heliocentric distance for two fields studied in this paper: vdB~130 field of
12 arcmin size around vdB~130 cluster (left panel) and protocluster field
(right panel). vdB~130 cluster members selected by CMD, distances and proper
motions criteria, listed in the Table~\ref{tab:new_vdB130}, are marked by
red, whereas field stars are marked by grey on the left panel on
Fig.~\ref{fig:Av}. It is clearly seen that the extinction systematically
grows with the distance up to 1.5~kpc, i.e. just to the distance of the
cluster, and remains nearly constant ($A_V \sim 2.5-3.0$~mag) at the
distances larger than 2~kpc, though with noticeable scatter. We can suppose
that this 2~kpc distance marks far edge of supershell around Cyg OB1
association. As for members of vdB~130 cluster, very large scatter of their
individual extinctions ($0 < A_V < 3$~mag) clearly seen on the left panel on
Fig.~\ref{fig:Av} undoubtedly confirms the presence of large differential
extinction inside this embedded young cluster, which is in qualitative
agreement with our data in the Table~\ref{tab:ages} for stars with spectral
classification (see also \citep{tatar16}).

Right panel on Fig.~\ref{fig:Av} shows the same dependence for protocluster area.
The picture shows nearly the same behavior of extinction with the distance along
line-of-sight as shown on Fig.~\ref{fig:ebv} (right panel) for our data on colour excess,
but with sharp increase of the extinction near $D \sim 1.5$~kpc.
Additionally, the comparison of two pictures imply nearly normal extinction law with
$R_V \sim 3$ \citep{Cardelli} toward protocluster direction.

\section{Conclusions.}

In this paper we investigate two sites of star formation in the wall
of the expanding supershell around the Cyg~OB1 association. Both
regions are associated with a cometary cloud, which is evidenced
not only by the factors mentioned in Papers I and II, but also
by our analysis of extinction. The region of ongoing star
formation -- the embedded cluster vdB~130 with an age $\leq 10$
Myr, containing 68 stars -- is located in the head of the cloud
`looking' toward the centre of the association. Whereas the
burst of star formation -- a compact protocluster containing at
least 30 type I and II protostars -- is observed in the tail of
the same cloud.

We analysed the optical components in the region using the data of the
GAIA~DR2 catalogue and our spectroscopic and \textit{BVRJHK} photometric
observations of stars in the region. We also used our original observations
to compare our data and data of the GAIA DR2 catalogue and obtained the
following results:

\begin{enumerate}

\item The use of high-precision GAIA parallaxes and proper motions
combined with spectroscopic observations and optical and IR photometry
allowed us to substantially refine the content and parameters of the open cluster vdB130.
The number of members of the vdB 130 cluster previously identified using UCAC4
catalogue increased as a result of new revision of astrometric data to 68 stars
with close proper motions (within 1 mas yr$^{-1}$) and trigonometric parallaxes
(lying in the interval from 0.50 to 0.70 mas). The cluster age (less than 10~Myr)
was estimated by the isochrone fitting technique applied to 8 stars with spectral
classes found from SED and colour-magnitude calibrations.

\item We show that the relative error of GAIA DR2 parallaxes
slightly increases with the distance to the object and strongly
depends on G magnitude. At the distance of 1.5--2~kpc it amounts
to 3--5 per cent (50--80~pc along line-of-sight) and to 20--30 per
cent (300--500~pc along line-of-sight) for bright and faint stars,
respectively.

\item The values of effective temperature provided by GAIA DR2s for stars located at
distances greater than $\sim$~0.8~kpc toward protocluster region are
systematically underestimated in comparison with the results of our optical
spectroscopic observations. This appears to be because of the method used to
determine  effective temperatures \citep{andrae} does not properly take into
account interstellar reddening.

\item The centroid of the proper motions of  vdB 130 members in
the Galactic coordinate system is located at $(pmL, pmB) \approx
(-6.20 \pm 0.33, +0.20 \pm 0.33)$ mas yr$^{-1}$. Hence the cluster
moves predominantly in the Galactic plane with a residual velocity
of about $-8$ km s$^{-1}$ directed toward the Galactic center
(after taking into account the differential rotation and solar
motion in accordance with the rotation curve of maser sources
(\citealt{ras17}).

\item We found no optical counterparts of the protocluster because
our analysis of 20 stars toward the protocluster direction show no clumping
in the distribution of proper motions or parallaxes. The stars studied are
distributed over a broad interval of heliocentric proper motions (these stars
span heliocentric distance from 0.7 to 3~kpc).

\item We studied the distance dependence of colour excess $E(B-V)$ and extinction $A_V$
for field stars and cluster members toward the vdB~130, and also for protocluster based
on our observations and data from StarHorse catalogue \citep{anders}.
The reddening is almost absent out to a distance of  $\approx 0.7-1$~kpc.
At larger distances toward the protocluster the extinction increase rate,
1.9 mag~kpc$^{-1}$ (determined assuming $R_V=3.08$), is approximately consistent
with the average rate in the Galaxy \citep{Sharov}.

\item toward vdB~130 cluster, the extinction measured for field stars systematically
increases up to 1.5~kpc, and after $\sim 2$~kpc remains nearly constant, with
$A_V \sim 2.5-3$~mag. The extinction increases much steeper than toward
protocluster possibly because we observe the stars through the dust and
gaseous layer of a supershell around Cyg~OB1. We do not exclude that the
distance 2~kpc marks far edge of the supershell. For cluster members, inside
the cluster large scatter of $A_V$ values from 0 to 3~mag confirms the idea
of large differential extinction.

\end{enumerate}

\section*{Data availability}
The data underlying this article will be shared on reasonable request to the corresponding author.

\section*{Acknowledgements}

The authors are grateful to Dr. Alexei Moiseev for carrying out spectroscopic
observations in July 2017, and to the reviewer, Dr. Alexander Binks, for his useful comments that improved our paper. 

This study was carried out using the equipment bought with the funds of the
Program of the Development of M.V. Lomonosov Moscow State University and
supported by the RFBR grants (18-02-00976, 18-02-00890, 19-12-00611, 20-02
-00643). Authors acknowledge the support from the Program of development of
M.V. Lomonosov Moscow State University (Leading Scientific School `Physics of
stars, relativistic objects and galaxies'). OE acknowledges the support by
Foundation of development of theoretical physics and mathematics `Basis'.

Observations with the Russian 6-m telescope carried out with the financial
support of the Ministry of Science and Higher Education of the Russian Federation.

This work is partially based on the data from the Spitzer Space
Telescope operated by the Jet Propulsion Laboratory, California
Institute of Technology, under NASA contract 1407, and the data
from the  2MASS catalogue (University of Massachusetts, California
Institute of Technology, NASA and NSF).

This work has made use of data from the European Space Agency
(ESA) mission {\it Gaia} (\url{https://www.cosmos.esa.int/gaia}),
processed by the {\it Gaia} Data Processing and Analysis
Consortium (DPAC,
\url{https://www.cosmos.esa.int/web/gaia/dpac/consortium}).

 \label{lastpage}
\end{document}